\documentclass{article}
\usepackage{amsmath,amsfonts,amssymb,setspace,cancel,url,hyperref,verbatim,color,todonotes}
\usepackage[a4paper]{geometry}
\usepackage{breqn,caption}
\usepackage{cite}
\newcommand{\be}{\begin{equation}}
\newcommand{\ee}{\end{equation}}

\newcommand{\ON}[1]{\mathrm{O}( #1 )}
\newcommand{\SU}[1]{\mathrm{SU}( #1 )}
\newcommand{\SL}[1]{\mathrm{SL}( #1 )}
\newcommand{\GL}[1]{\mathrm{GL}( #1 )}

\newcommand{\Spin}[1]{\mathrm{Spin}(#1)}

\newcommand{\USp}[1]{\mathrm{USp}(#1)}

\newcommand{\TA}{{\cal A}}
\newcommand{\TB}{{\cal B}}
\newcommand{\TC}{{\cal C}}
\newcommand{\TD}{{\cal D}}

\newcommand{\Fa}{\mathcal{F}}
\newcommand{\Fb}{\mathcal{H}}
\newcommand{\Fc}{\mathcal{J}}

\newcommand{\D}{\mathfrak{D}}

\newcommand{\hpartial}{\hat{\partial}}

\newcommand{\bmu}{\bar{\mu}}

\newcommand{\bi}{\bar{i}}

\newcommand{\tomega}{\tilde{\omega}}

\newcommand{\dalpha}{\dot{\alpha}}
\newcommand{\dbeta}{\dot{\beta}}
\newcommand{\dgamma}{\dot{\gamma}}

\newcommand{\drho}{\dot{\rho}}

\newcommand{\ctheta}{\left(\theta^*\right)}

\newcommand{\tnabla}{\tilde{\nabla}}
\newcommand{\tGamma}{\tilde{\Gamma}}
\newcommand{\obf}[1]{\overline{\mathbf{#1}}}
\newcommand{\mbf}[1]{\mathbf{#1}}
\newcommand{\gL}{\mathcal{L}}

\newcommand{\gM}{\mathcal{M}}
\newcommand{\gH}{\mathcal{H}}
\newcommand{\cR}{\mathcal{R}}

\newcommand{\Vi}{{\cal V}}
\newcommand{\tV}{\tilde{V}}
\newcommand{\bae}{\bar{e}}
\newcommand{\bag}{\bar{g}}
\numberwithin{equation}{section}

\newcommand\Tstrut{\rule{0pt}{3ex}}         
\newcommand\Bstrut{\rule[-1.3ex]{0pt}{0pt}}   

\begin{document}

\begin{titlepage}
\vfill

\begin{flushright}
LMU-ASC 61/16
\end{flushright}

\vfill

\begin{center}
   \baselineskip=16pt
   	{\Large \bf 7-dimensional ${\cal N}=2$ Consistent Truncations using $\SL{5}$ Exceptional Field Theory}
   	\vskip 2cm
 	{\large \bf Emanuel Malek} 
   	\vskip .6cm
   	{\it Arnold Sommerfeld Center for Theoretical Physics, Department f\"ur Physik, \\ Ludwig-Maximilians-Universit\"at M\"unchen, Theresienstra{\ss}e 37, 80333 M\"unchen, Germany}
   	\vskip .6cm
   	{{E.Malek@lmu.de}} \\
   	\vskip 2cm
\end{center}
\vfill

\begin{abstract}
We show how to construct seven-dimensional half-maximally supersymmetric consistent truncations of 11-/10-dimensional SUGRA using $\SL{5}$ exceptional field theory. Such truncations are defined on generalised $\SU{2}$-structure manifolds and give rise to seven-dimensional half-maximal gauged supergravities coupled to $n$ vector multiplets and thus with scalar coset space $\mathbb{R}^+ \times \ON{3,n}/\ON{3}\times\ON{n}$. The consistency conditions for the truncation can be written in terms of the generalised Lie derivative and take a simple geometric form. We show that after imposing certain ``doublet'' and ``closure'' conditions, the embedding tensor of the gauged supergravity is given by the intrinsic torsion of generalised $\SU{2}$-connections, which for consistency must be constant, and automatically satisfies the linear constraint of seven-dimensional half-maximal gauged supergravities, as well as the quadratic constraint when the section condition is satisfied.
\end{abstract}

\vfill

\setcounter{footnote}{0}
\end{titlepage}

\tableofcontents

\newpage

\section{Introduction} \label{s:Introduction}
Finding consistent truncations of higher-dimensional supergravity to yield lower-dimensional theories is a notoriously difficult problem. By a consistent truncation we mean that solutions of the lower-dimensional equations of motion automatically satisfy those of the initial higher-dimensional theory. Because of the non-linearity of the field equations such consistent truncation Ans\"atze are generically hard to find \cite{Duff:1984hn}, unless the background has a lot of underlying symmetry. For example, Scherk and Schwarz \cite{Scherk:1979zr} showed that consistent truncations can be defined on Lie groups, which are of course parallelisable manifolds. As a result the truncation has the same number of supersymmetries as the higher-dimensional theory.

Recently, the Scherk-Schwarz set-up has been generalised using double field theory (DFT) \cite{Siegel:1993th,Siegel:1993xq,Hull:2006va,Hull:2009mi} and exceptional field theory (EFT) \cite{Hull:2007zu,Berman:2010is,Hohm:2013uia}, as well as generalised geometry \cite{Gualtieri:2003dx,Hitchin:2004ut,Coimbra:2011nw,Coimbra:2011ky}. These theories are $\ON{D,D}$- and $E_{d(d)}$-manifest extensions (or reformulations in the case of generalised geometry) of 10-/11-dimensional supergravity which treat the gauge and gravitational fields on an equal footing, see \cite{Duff:1989tf,Duff:1990hn,Tseytlin:1990nb,Tseytlin:1990va} for earlier work in this direction. They thus naturally include fluxes in the Scherk-Schwarz set-up. As a result, a generalised Scherk-Schwarz Ansatz \cite{Geissbuhler:2011mx,Grana:2012rr,Dibitetto:2012rk,Geissbuhler:2013uka,Berman:2012uy,Hohm:2014qga,Coimbra:2012af,Godazgar:2013dma,Godazgar:2013oba,Godazgar:2013pfa,Lee:2014mla,Lee:2015xga} can be performed on a background which is ``generalised parallelisable'' \cite{Lee:2014mla}, the flux-analogue of a parallelisable manifold. Such backgrounds may not be parallelisable as a differential manifold and indeed one can show that an otherwise remarkable set of consistent truncations on spheres, in particular $S^7$ \cite{deWit:1986mz} and $S^4$ \cite{Nastase:1999cb,Nastase:1999kf} of 11-dimensional SUGRA and $S^5$ \cite{Cvetic:2000nc} for IIB SUGRA, can be understood as such generalised Scherk-Schwarz Ans\"atze \cite{Lee:2014mla,Hohm:2014qga}.

With this set-up it has been possible to derive and study a variety of new consistent truncations on spheres and hyperboloids \cite{Hohm:2014qga,Baguet:2015sma,Cassani:2016ncu}, on non-geometric backgrounds \cite{Dibitetto:2012rk,Hassler:2014sba}, on product manifolds \cite{Baguet:2015iou} as well as to study the relationship between different consistent truncations \cite{Malek:2015hma,Malek:2015etj}.\footnote{The closely-related approach of \cite{Ciceri:2014wya,Guarino:2015jca,Guarino:2015vca} has also been fruitful in finding consistent truncations.} Because such generalised Scherk-Schwarz truncations are defined on generalised parallelisable spaces, they also preserve the same number of supersymmetries \cite{Coimbra:2011ky} and thus yield maximal gauged SUGRAs when used in EFT or half-maximal gauged SUGRAs when used in DFT, or their respective generalised geometry analogues. While it is possible to define a further truncation of the maximal gauged SUGRA to half-maximal ones, corresponding to the reduction of EFT to DFT, see e.g. \cite{Dibitetto:2012rk,Dibitetto:2015bia}, there are of course half-maximal gauged SUGRAs which cannot be obtained this way. Furthermore, there are half-maximal gauged SUGRAs which cannot be obtained by a consistent truncation of type II theories but require either the heterotic SUGRA or 11-dimensional SUGRA.

The purpose of this paper is to initiate the study of consistent truncations which break some amount of supersymmetry in exceptional field theory (and exceptional generalised geometry). Here we will focus on the seven-dimensional case where we show how to obtain arbitrary half-maximal gauged SUGRAs coupled to $n$ vector multiplets and thus with scalar coset space
\begin{equation}
 M_{scalar} = \frac{\ON{3,n}}{\ON{3}\times\ON{n}} \times \mathbb{R}^+ \,.
\end{equation}
Our set-up allows us to capture consistent truncations to half-maximal gauged SUGRAs arising from either 11-dimensional or type II SUGRA, as well as the heterotic SUGRA, as shown in \cite{Malek:2016vsh}. \footnote{We should mention that the approach we take here differs from that in \cite{Blumenhagen:2015lta} which reduces the usual flux formulation of double field theory, which is only valid for paralellisable manifolds, on $CY_3$ to obtain a ${\cal N}=2$ scalar potential. Furthermore, our approach allows us to consider general flux backgrounds whereas \cite{Blumenhagen:2015lta} is restricted to fluxes which can be treated as small deviations to the Calabi-Yau background.}

In order to break half of the supersymmetry, the internal manifold must have generalised $\SU{2}$-structure and we show how to define consistent truncations on such spaces. In particular, the embedding tensor is encoded in the generalised Lie derivative acting on the sections defining the truncation and automatically satisfies the linear constraint of half-maximal gauged SUGRA. Exactly as in the maximal case \cite{Grana:2012rr}, the section condition is sufficient for the gaugings to fulfil the quadratic constraint.

In order to understand how to obtain $n\neq3$ vector multiplets it is important to distinguish between linear symmetry groups acting at each point in space and the symmetry group acting on the sections we keep in the truncation. The linear symmetry group is in the case of exceptional field theory just $E_{d(d)}$, or in the case considered here $\SL{5}$. This is simply a consequence of the field content of the theory, and not a result of the backgrounds considered.\footnote{In this discussion we ignore the existence of the extra coordinates but we show how these fit into the picture in \cite{Malek:2016vsh}.} However, when we consider truncations on generalised parallelisable spaces then this also becomes the symmetry group acting on sections and this is why generalised Scherk-Schwarz reductions lead to gauged SUGRAs with global symmetry group $E_{d(d)}$. On the other hand, when the background is not generalised parallelisable, as we will be considering in this paper, the group acting on the space of sections can be much larger because the number of independent sections can be larger. This is why the supergravities we obtain have global symmetry groups $\ON{3,n}$ which are clearly not subgroups of $\SL{5}$.

To emphasise this point, let us consider the more familiar example of general relativity in $d+4$-dimensions on a product manifold so that its linear symmetry group is $\GL{d}\times\GL{4}$. When performing a truncation on $T^4$, one obtains $d$-dimensional gravity minimally coupled to scalars parameterising a coset whose global symmetry group is indeed $\GL{4}$. However, when considering less supersymmetric truncations, for example on K3, one obtains duality groups which are not subgroups of $\GL{4}$. In the K3 example one obtains $d$-dimensional gravity minimally coupled to scalars parameterising the coset space $\ON{3,19}/\ON{3}\times \ON{9}$. The duality group $\ON{3,19}$ acts of course on the space of sections defining the truncation on K3, i.e. the 22 harmonic forms. The linear symmetry group of the internal space, $\GL{4}$, which just tells us that at each point we have a $4$-dimensional metric plays no direct role in the global symmetry group of the reduced theory, $\ON{3,19}$.

We begin by reviewing the $\SL{5}$ EFT relevant for truncations to 7-dimensional gauged SUGRAs in section \ref{s:Overview} and introducing the tensors required to define a $\SU{2}$-structure in section \ref{s:SpinorBilinears}. Then we reformulate the theory in section \ref{s:Reformulate} in a way that is more adapted to ${\cal N}=2$ SUSY. This involves rewriting the theory in terms of tensors defining the $\SU{2}$-structure group rather than the generalised metric. That such a reformulation bypassing the generalised metric is necessary should not come as a surprise since as a particular example one could here consider the M-theory truncation on K3 for which the metric, and hence generalised metric, is not explicitly known. We show how to rewrite the supersymmetry variation of the gravitino as well as the scalar potential, kinetic terms and topological term in a way that is adapted to ${\cal N}=2$ SUSY.

We next discuss how to perform a truncation of EFT on generalised $\SU{2}$-structure manifolds in section \ref{s:Truncation}. The conditions for a consistent truncation are compactly formulated in terms of the generalised Lie derivative. In particular, with the truncation Ansatz we present the internal coordinates can only appear in the action through the embedding tensor, which is defined by the generalised Lie derivative of the sections defining the truncation. Thus when the embedding tensor components are constant, the Ansatz guarantees that the action becomes independent of the internal coordinates and thus the truncation is consistent. Finally, we conclude in section \ref{s:Conclusions} by discussing possible application and extensions of this work.

\paragraph{Summary of results}
Throughout this paper we are concerned with generalised $\SU{2}$-structure manifolds. Such manifolds admit two linearly-independent, nowhere-vanishing spinors. This is equivalent to the manifold having the following nowhere-vanishing tensors under generalised diffeomorphisms: $\left(\kappa, A_a, A^a, B_{u,ab}\right)$. Here $a, b = 1, \ldots, 5$ denote $\SL{5}$ indices and $u = 1, \ldots, 3$ are $\SU{2}_R$ indices denoting the R-symmetry. Furthermore, $\kappa$ is a scalar density of weight $\frac15$ and is related to the determinant of the external seven-dimensional metric. Additionally these structures satisfy
\begin{equation}
 A^a A_a = \frac{1}{2} \,, \qquad B_{u,ab} A^a = 0 \,, \qquad \epsilon^{abcde} B_{u,ab} B_{v,cd} = 4 \sqrt{2} A^e \,.
\end{equation}
This set of tensors reduce the $\USp{4}$-structure group to $\SU{2}$ and thus define a generalised metric implicitly.

The action can be rewritten completely in terms of the generalised $\SU{2}$-structure, i.e. $\kappa$, $A_a$, $A^a$, $B_{u,ab}$. To do so one introduces a generalised $\SU{2}$-connection $\tnabla$ which annihilates the $\SU{2}$-structure
\begin{equation}
 \tnabla_{ab} \kappa = \tnabla_{ab} A_c = \tnabla_{ab} A^c = \tnabla_{ab} B_{u,ab} = 0 \,.
\end{equation}
Its intrinsic torsion lives in the representations
\begin{equation}
 W_{int} = 2 \cdot \left( \mbf{1},\mbf{1} \right) + 2 \cdot \left(\mbf{1},\mbf{3}\right) + \left(\mbf{3},\mbf{1}\right) + \left(\mbf{3},\mbf{3}\right) + 3 \cdot \left(\mbf{2},\mbf{2}\right) + \left(\mbf{2},\mbf{4}\right) \,,
\end{equation}
of $\SU{2}_S \times \SU{2}_R \subset \SL{5}$, where $\SU{2}_S$ refers to the structure group while $\SU{2}_R$ refers to the R-symmetry group. The intrinsic torsion can be used to rewrite the SUSY variations and the scalar potential. For example, the generalised Ricci scalar is given by
\begin{equation}
 \begin{split}
  \cR &= 8\, S^2 - 2\, T^2 - 8\sqrt{2}\, ST - 3\, T_u T^u + T_u S^u - \frac34\, S_u S^u - 16\sqrt{2}\, \epsilon^{abcde} T_{ab} T_{cd} A_e \\
  & \quad - 36\sqrt{2}\, \epsilon^{abcde} T^u{}_{ab} T_{u,cd} A_e - \frac{4\sqrt{2}}{3}\, M^{ab} S_a S_b - \frac{16}{3}\, M^{ab} S_a T_b + \frac83\, M^{ab} U_a S_b \,.
 \end{split}
\end{equation}
where $S$, $T$ are singlets of the intrinsic torsion, $T_u$ $S_u$ are $\left(\mbf{1},\mbf{3}\right)$ under $\SU{2}_S\times \SU{2}_R$, $T_{ab}$ are $\left(\mbf{3},\mbf{1}\right)$ under $\SU{2}_S\times \SU{2}_R$, $T_u{}^{ab}$ are $\left(\mbf{3},\mbf{3}\right)$ and $S_a$, $T_a$, $U_a$ are the $\left(2,2\right)$ of the intrinsic torsion.

We perform a truncation by expanding the $\SU{2}$-structure in terms of a finite basis of sections of the $\left(\mbf{1},\mbf{1}\right)$, $\left(\mbf{1},\mbf{3}\right)$ and $\left(\mbf{3},\mbf{1}\right)$-bundles of $\SU{2}_S\times \SU{2}_R$. In particular because $\SU{2}_S$ is non-trivially fibred over the manifold we use $n$ sections of the $\left(\mbf{3},\mbf{1}\right)$-bundle and these will give rise to $n$ vector multiplets in the reduced theory. We denote the sections by $n^a$, $n_a$ and $\omega_{M,ab}$, where $M = 1, \ldots n+3$ collective denotes the sections of the $\left(\mbf{3},\mbf{1}\right)$ and $\left(\mbf{1},\mbf{3}\right)$-bundles. These sections satisfy
\begin{equation}
 n^a n_a = 1 \,, \qquad \omega_{M,ab} n^a = 0 \,, \qquad \omega_{M,ab} \omega_{N,cd} \epsilon^{abcde} = 4 \eta_{MN} n^e \,,
\end{equation}
where $\eta_{MN}$ is an $\ON{3,n}$ metric whose signature depends on the number of $\left(\mbf{3},\mbf{1}\right)$ sections.

The truncation Ansatz for the scalars is given by
\begin{equation}
 \begin{split}
  \langle\kappa\rangle(x,Y) &= |\bae|^{1/7}(x)\,e^{-2d(x)/5}\, \rho(Y) \,, \\
  \langle A^a\rangle(x,Y) &= \frac{1}{\sqrt{2}} e^{-4d(x)/5} n^a(Y) \,, \\
  \langle A_a\rangle(x,Y) &= \frac{1}{\sqrt{2}} e^{4d(x)/5} n_a(Y) \,, \\
  \langle B_{u,ab} \rangle(x,Y) &= e^{-2d(x)/5}\, b_{u,M}(x) \omega^M{}_{ab}(Y) \,,
 \end{split}
\end{equation}
where we use $\langle \, \rangle$ to denote the truncation Ansatz, and $\rho(Y)$ is a density of weight $\frac15$ under generalised diffeomorphisms. The scalars $b_u{}^M$ then satisfy
\begin{equation}
 b_u{}^M b_{v,M} = \delta_{uv}
\end{equation}
and parameterise the coset $\frac{\ON{3,n}}{\ON{3}\times\ON{n}}$. Similarly $|\bae|$ and $d(x)$ are the determinant of the 7-dimensional metric and the dilaton, respectively.

In order to have a consistent truncation, the sections $\rho$, $n^a$, $n_a$ and $\omega_{M,ab}$ must satisfy three types of differential constraints. Firstly, any doublets must vanish, e.g.
\begin{equation}
 n^a \gL_{\tomega_{M}} \tomega_{M}{}^{ab} = 0 \,,
\end{equation}
where we defined the $n+3$ generalised vectors
\begin{equation}
 \tomega_M{}^{ab} = \rho\, \omega_M{}^{ab} \,, \qquad \textrm{with} \quad \omega_M{}^{ab} = \epsilon^{abcde} \omega_{M,cd} n_e \,.
\end{equation}
Secondly, the generalised Lie derivative of the sections $\omega_{M,ab}$ must be expandable in a basis of the $\omega_{M,ab}$.

The embedding tensor of the half-maximal gauged supergravity is then given by the generalised Lie derivative of the sections defining the truncation. In particular, this satisfies the linear constraint of 7-d half-maximal gauged supergravities so that one can identify
\begin{equation}
 \begin{split}
  f_{MNP} &= \frac1{4\rho} \gL_{\tomega_{[M}} \omega_{N|ab|} \omega_{P]}{}^{ab} \,,\\
  f_M &= n^a \gL_{\tomega_M} n_a \,, \qquad \xi_M = \rho^{-1} \gL_{\tomega_M} \rho \,,\\
  \Theta &= \rho n^a \partial_{ab} n^b \,.
 \end{split} \label{eq:EmbeddingTensorSummary}
\end{equation}
By construction, closure of the algebra of generalised Lie derivatives (hence for example the section condition) is sufficient for the gaugings to satisfy the quadratic constraints of the gauged SUGRAs. Finally, the truncation is consistent when the embedding tensor \eqref{eq:EmbeddingTensorSummary} is constant.

Unlike in the construction of effective actions, the $\omega_{M,ab}$'s appearing here are not uniquely defined by the topology of the background. This is a reflection of the fact that a given background can admit multiple, different consistent truncations. Additionally, it is important to highlight that the consistent truncations defined here do not require the background to be a solution of the equations of motion. In this case, the gauged SUGRA will not have a vacuum at the origin of the scalar manifold, nor does it need to have a vacuum at all. Related to this, the fields in the truncated theory are not in general massless. In particular, the consistent truncation may have discarded some light modes but kept certain heavier modes. However, it does so in a manner in which any solutions can be uplifted to solutions of the full theory.

\section{Overview of exceptional field theory} \label{s:Overview}
Let us begin by giving a brief review of the $\SL{5}$ exceptional field theory \cite{Berman:2010is,Hohm:2013uia,Musaev:2015ces} with emphasis on the aspects needed for our purposes. We refer the interested reader to the reviews \cite{Berman:2013eva,Aldazabal:2013sca,Hohm:2013bwa}. The $\SL{5}$ EFT can be viewed as a reformulation of 11-dimensional supergravity which makes the linear symmetry group $\SL{5}$ manifest. Thus, the starting point is 11-dimensional supergravity in a 7+4 split. Let us use $x^\mu$, $\mu = 1, \ldots, 7$, as coordinates for the ``external'' 7-d space and label $y^{\bar{i}}$, $\bi = 1, \ldots, 4$ as the four ``internal coordinates''. These are part of 10 ``extended coordinates'', $Y^{ab}$, forming the antisymmetric representation of $\SL{5}$, where we use $a, b = 1, \ldots, 5$ as fundamental $\SL{5}$ indices. In the case where the internal geometry really is a torus, the extra six coordinates can be understood as being dual to wrapping modes of branes. However, the extra coordinates are always introduced, in a background-independent manner, and we will suggest a possible interpretation in the case where the four-dimensional part of the internal space is non-toroidal, e.g. a K3, in \cite{Malek:2016vsh}.\footnote{In the case of double field theory this process is a little bit clearer. There one doubles the ``internal'' space, corresponding to independent zero modes of left- and right-movers which one could introduce for a string propagating in an arbitrary background. In the case of a toroidal background these zero modes are indeed dual to momentum and winding modes of the string. In EFT an analogous zero-mode interpretation is lacking.} We will always refer to the seven-dimensional space as external and the four-dimensional (or 10-dimensional if the extended viewpoint is taken) as ``internal'' although no truncation has been performed, i.e. all fields can depend on any of the $\left(7+10\right)$ coordinates.

All scalars with respect to this $\left(7+4\right)$-split can be described by the generalised metric
\begin{equation}
 \gM_{ab} \in \SL{5} / \USp{4}.
\end{equation}
This coset can also be described by the generalised vielbein $\Vi_a{}^{ij}$ such that
\begin{equation}
 \gM_{ab} = \Vi_a{}^{ij} \Vi_{b,ij} \,,
\end{equation}
where $i, j = 1, \ldots, 4$ are $\USp{4}$ indices which are raised/lowered by the symplectic invariant $\Omega_{ij}$. The $\Vi_a{}^{ij}$ furthermore satisfy
\begin{equation}
 \Vi_a{}^{(ij)} = 0 \,, \qquad \Vi_a{}^{ij} \Omega_{ij} = 0 \,, \qquad \left( \Vi_a{}^{ij} \right)^* = \Vi_{a,ij} \,.
\end{equation}
See \cite{Samtleben:2005bp} for more $\USp{4}$ conventions which we here largely follow. Similarly, all bosonic objects with one leg in the external space can be combined into 10 vector fields $\TA_\mu{}^{ab}$. Those with two external legs can be combined into five two-forms $\TB_{\mu\nu,a}$, etc.

Just as the bosonic degrees of freedom form $\SL{5}$ representations, so too do the local symmetries of 11-dimensional supergravity, i.e. diffeomorphisms and $p$-form transformations. The symmetries acting on the internal space combine into so-called generalised diffeomorphisms generated by the generalised Lie derivative. For a tensor in the $\SL{5}$ fundamental representation $V^a$ of weight $\frac15$ this takes the form \cite{Berman:2011cg,Coimbra:2011ky,Berman:2012vc}
\begin{equation}
 \gL_\Lambda V^a = \frac12 \Lambda^{bc} \partial_{bc} V^a - V^b \partial_{bc} \Lambda^{ac} + \frac15 V^a \partial_{bc} \Lambda^{bc} +\frac{\lambda}{2} V^a \partial_{bc} \Lambda^{bc} \,, \label{eq:LieDerivative}
\end{equation}
and for a scalar
\begin{equation}
 \gL_{\Lambda} S = \frac12 \Lambda^{ab} \partial_{ab} S \,.
\end{equation}
All other cases follow by linearity. Note that from the above considerations $\partial_{ab}$ can be seen to carry weight $-\frac15$ under generalised diffeomorphisms. Furthermore, the parameter of generalised diffeomorphisms $\Lambda^{ab}$ is in the $\mathbf{10}$ of $\SL{5}$ and has weight $\frac15$, so that under a generalised diffeomorphism it itself transforms as
\begin{equation}
 \gL_{\Lambda_1} \Lambda_2^{ab} = \frac12 {\Lambda_1}^{cd} \partial_{cd} \Lambda_2^{ab} + \left( \frac25 + \frac{1}{10} \right) \Lambda_2^{ab} \partial_{cd} \Lambda_1^{cd} - \Lambda_2^{cb} \partial_{cd} \Lambda_1^{ad} - \Lambda_2^{ac} \partial_{cd} \Lambda_1^{bd} \,.
\end{equation}
We will henceforth call any tensors in the $\mathbf{10}$ of $\SL{5}$ of weight $\frac15$ ``generalised vectors'', because they generate generalised diffeomorphisms.

For consistency the algebra of generalised diffeomorphisms must close, i.e.
\begin{equation}
 \left[ \gL_{\Lambda_1}, \gL_{\Lambda_2} \right] V^a = \gL_{\left[\Lambda_1,\Lambda_2\right]_D} V^a \,. \label{eq:AlgebraClosure}
\end{equation}
Here the $D$-bracket just represents the action of a generalised Lie derivative,
\begin{equation}
 \left[ \Lambda_1, \Lambda_2 \right]_D^{ab} = \gL_{\Lambda_1} \Lambda_2^{ab} \,.
\end{equation}
In order for \eqref{eq:AlgebraClosure} to hold one needs to impose the so-called section condition\cite{Berman:2011cg,Coimbra:2011ky}
\begin{equation}
 \partial_{[ab} f \partial_{cd]} g = 0 \,, \qquad \partial_{[ab} \partial_{cd]} f = 0 \,,
\end{equation}
where $f$ and $g$ denote any two objects of the $\SL{5}$ EFT. There are two inequivalent solutions to the section condition, one corresponding to 11-dimensional SUGRA while the other corresponds to type IIB \cite{Blair:2013gqa,Hohm:2013uia,Hohm:2013vpa}. Upon using a solution of the section condition, the generalised Lie derivative \eqref{eq:LieDerivative} generates the $p$-form gauge transformation and diffeomorphisms of the corresponding SUGRA. Similarly, the action that we are about to sketch reduces to the 11-dimensional SUGRA or IIB SUGRA action, upon imposing a solution of the section condition. However, one could also consider a set-up where there is not a globally well-defined solution to the section condition, in which case we obtain a non-geometric background.

Given the generalised Lie derivative, one can introduce connections which give covariant derivatives with respect to these generalised diffeomorphisms. As usual one can also introduce a torsion as the tensorial part of a connection $\nabla$. This can be conveniently defined via the generalised Lie derivative as
\begin{equation}
 \gL_{\Lambda}^\nabla V^a - \gL_\Lambda V^a = \frac12 \tau_{bc,d}{}^a \Lambda^{bc} V^d + \frac{\lambda}{2} \tau_{bc} \Lambda^{bc} V^d \,, \label{eq:LieTorsion}
\end{equation}
where $\gL_{\Lambda}^\nabla$ denotes the generalised Lie derivative \eqref{eq:LieDerivative} with all partial derivatives replaced by the covariant derivatives $\nabla_{ab}$. It can be shown \cite{Coimbra:2011ky,Cederwall:2013naa,Blair:2014zba} that the torsion lives in the following irreps of $\SL{5}$
\begin{equation}
 \tau_{ab,c}{}^d \in \mathbf{10} \oplus \mathbf{15} \oplus \obf{40} \,.
\end{equation}

Using these concepts one can, for example introduce a generalised torsion-free $\USp{4}$ connection \cite{Coimbra:2011ky,Coimbra:2012af,Hohm:2013uia,Cederwall:2013naa}. This connection is particularly useful for coupling fermions \cite{Godazgar:2014nqa,Musaev:2014lna,Baguet:2016jph} and can also be used to derive a ``generalised curvature scalar''. We will make use of it throughout this paper and label it by $\nabla_{ab}$. However, it is important to note that the torsion constraint does not fix the connection uniquely. Instead, only certain irreducible representations are uniquely fixed, see e.g. \cite{Coimbra:2011ky,Coimbra:2012af}. The generalised curvature scalar that can be derived in this way is in fact a scalar density under generalised diffeomorphisms which only involves derivatives with respect to the internal space of $\gM_{ab}$ and $g_{\mu\nu}$. It is the EFT lift of the scalar potential of seven-dimensional gauged SUGRAs. Conversely, it reduces to the scalar potential of maximal seven-dimensional gauged SUGRAs upon imposing a Scherk-Schwarz Ansatz \cite{Berman:2012uy,Hohm:2014qga}. We should mention that there are also other geometric ways of constructing the generalised curvature scalar, e.g. \cite{Park:2013gaj,Blair:2014zba}.

In order to define the EFT on the full $\left(7+10\right)$-dimensional space one needs to introduce a seven-dimensional derivative which is covariant under generalised diffeomorphisms. This is given by the covariant external derivative
\begin{equation}
 D_\mu = \partial_\mu - \gL_{A_\mu} \,, \label{eq:CovExtDeriv}
\end{equation}
and upon Scherk-Schwarz reduction this reduces to the gauge-covariant derivative of the gauged SUGRA.

The final ingredient required for constructing the EFT action are the field strengths of the vector fields, two-form and three-form potentials. We will label these as $\TA_\mu{}^{ab}$, $\TB_{\mu\nu,a}$, $\TC_{\mu\nu\rho}{}^a$ and $\TD_{\mu\nu\rho\sigma,ab}$, which is the auxiliary 4-form potential appearing in the action without kinetic term \cite{Hohm:2013uia,Wang:2015hca,Hohm:2015xna}. These have weights $\frac15$, $\frac25$, $\frac35$ and $\frac45$, respectively, under generalised diffeomorphisms. Following \cite{Wang:2015hca,Hohm:2015xna}, their field strengths can be written in $\SL{5}$ index-free notation as
\begin{equation}
\begin{split}
\Fa_{\mu\nu} &= 2\partial_{[\mu}\TA_{\nu]} - [\TA_{\mu},\TA_{\nu}]_E + \hpartial \TB_{\mu\nu} \,, \\
\Fb_{\mu\nu\rho} &= 3\D_{[\mu}\TB_{\nu\rho]} - 3\partial_{[\mu}\TA_{\nu}\bullet\TA_{\rho]} + \TA_{[\mu}\bullet[\TA_{\nu},\TA_{\rho]}]_E + \hpartial\TC_{\mu\nu\rho} \,, \\
\Fc_{\mu\nu\rho\sigma} &= 4\D_{[\mu}\TC_{\nu\rho\sigma]} + 3\hpartial\TB_{[\mu\nu}\bullet\TB_{\rho\sigma]} - 6\Fa_{[\mu\nu}\bullet\TB_{\rho\sigma]} + 4\TA_{[\mu}\bullet(\TA_{\nu}\bullet\partial_{\rho}\TA_{\sigma]}) \\
& \quad - \TA_{[\mu}\bullet(\TA_{\nu}\bullet[\TA_{\rho},\TA_{\sigma]}]_E) + \hpartial\TD_{\mu\nu\rho\sigma} \,,
\end{split}
\label{eq:fieldstrengths}
\end{equation}
where the $E$-bracket is the antisymmetrised generalised Lie derivative
\begin{equation}
 \left[ V, W \right]_E = \frac12 \left( \gL_V W - \gL_W V \right) \,,
\end{equation}
the $\bullet$ operation is defined as
\begin{equation}
 \begin{split}
  \left(\TA_1 \bullet \TA_2\right)_a &= \frac14 \epsilon_{abcde} \TA_1^{bc} \TA_2^{de} \,, \\
  \left(\TA \bullet \TB\right)^a &= \TA^{ab} \TB_{b} \,, \\
  \left(\TA \bullet \TC\right)_{ab} &= \frac14 \epsilon_{abcde} \TA^{cd} \TC^e \,, \\
  \TA \bullet \TD &= \frac12 \TA^{ab} \TD_{ab} \,, \\
  \left( \TB_1 \bullet \TB_2 \right)_{ab} &= \TB_{2[a} \TB_{|1|b]} \,, \\
  \TB \bullet \TC &= \TB_a \TC^a \,, \label{eq:Bullet}
 \end{split}
\end{equation}
and the (nilpotent) derivative $\hpartial$ is
\begin{equation}
 \hat{\partial} \TB^{ab} = \frac{1}{2} \epsilon^{abcde} \partial_{cd} \TB_e \,, \qquad \hat{\partial} \TC_a = \partial_{ba} \TC^b \,, \qquad \hat{\partial} \TD^a = \frac12 \epsilon^{abcde} \partial_{bc} \TD_{de} \,. \label{eq:NilPot}
\end{equation}
Note that the derivative $\hpartial$ is a covariant derivative when acting on objects with the appropriate weight, i.e. when $\TB_a$ has weight $\frac25$, $\TC^a$ has weight $\frac35$ and $\TD_{ab}$ has weight $\frac45$.

With all these ingredients one can construct the $\SL{5}$ EFT action \cite{Hohm:2013uia,Musaev:2015ces,Bosque:2016fpi} as
\begin{equation}
 S = \int d^{10}Y d^7x |e| \left( L_{EH} + L_{SK} + L_{GK} - V \right) + S_{top} \,. \label{eq:action}
\end{equation}
Here $L_{EH}$ is the seven-dimensional modified Einstein-Hilbert term, where all $\partial_\mu$ are replaced by $D_\mu$ \cite{Berman:2015rcc}, in order to be invariant under generalised diffeomorphisms. This is necessary because the seven-dimensional metric $g_{\mu\nu}$ is not a scalar but a density of weight $\frac25$ under generalised diffeomorphisms. The alternative is to use the vielbein formalism \cite{Hohm:2013vpa}. We define the modified Riemann tensor as
\begin{equation}
 R^{\mu}{}_{\nu\rho\sigma} = \D_\rho \Gamma^\mu{}_{\nu\sigma} - \D_\sigma \Gamma^\mu{}_{\nu\rho} + \Gamma^\mu{}_{\lambda\rho} \Gamma^\lambda{}_{\nu\sigma} - \Gamma^\mu{}_{\lambda\sigma} \Gamma^\lambda{}_{\nu\rho} \,,
\end{equation}
where
\begin{equation}
 \Gamma^\mu{}_{\nu\rho} = g^{\mu\sigma} \left( \D_{(\nu} g_{\rho)\sigma} - \frac12 \D_\sigma g_{\nu\rho} \right) \,.
\end{equation}
The modified Einstein-Hilbert term is then
\begin{equation}
 L_{EH} = g^{\mu\nu} R^{\rho}{}_{\mu\rho\nu} \,.
\end{equation}

Furthermore,
\begin{equation}
 \begin{split}
  L_{SK} &= \frac14 g^{\mu\nu} D_\mu \gM^{ab} D_\nu \gM_{ab} \,, \\
  L_{GK} &= -\frac18 \left( \Fa_{\mu\nu}{}^{ab} \Fa^{\mu\nu,cd} \gM_{ac} \gM_{bd} + \frac{2}{3} \Fb_{\mu\nu\rho,a} \Fb^{\mu\nu\rho}{}_{b} \gM^{ab} \right) \,, \\
  V &= - \left( \frac14\cR + \frac18 \gM^{ac} \gM^{bd} \nabla_{ab} g_{\mu\nu} \nabla_{cd} g^{\mu\nu} \right) \,,
 \end{split}
\end{equation}
where $\cR$ is the generalised Ricci scalar \cite{Coimbra:2012af,Hohm:2013uia} which involves only internal derivatives of the generalised metric. The topological term is best written as an integral over a 10-dimensional extended space and an eight-dimensional external spacetime, whose boundary is the seven-dimensional external spacetime we are considering \cite{Hohm:2015xna,Wang:2015hca,Musaev:2015ces,Bosque:2016fpi}
\begin{equation}
 S_{top} = - \frac{1}{2\sqrt{6}} \int \mathrm{d}^{10}Y\, \mathrm{d}^8x \left( \frac14 \hpartial \Fc_{\mu_1\ldots\mu_4} \bullet \Fc_{\mu_5\ldots\mu_8} - 4 \Fa_{\mu_1\mu_2} \bullet \left( \Fb_{\mu_3\ldots\mu_5} \bullet \Fb_{\mu_6\ldots\mu_8} \right) \right) \epsilon^{\mu_1\ldots\mu_8} \,. \label{eq:TopTerm}
\end{equation}
While each of these terms is individually a scalar (density) under generalised diffeomorphisms, it transforms anomalously under external diffeomorphisms. The various coefficients are fixed uniquely in order to ensure that the entire Lagrangian is invariant under external spacetime diffeomorphisms.

\section{Spinor bilinears and $\SU{2}$ structure} \label{s:SpinorBilinears}
In order to obtain a half-maximal theory in seven-dimensions, the internal space must admit two globally well-defined spinors, $\Theta_1$ and $\Theta_2$. These two spinors form a $\SU{2}_R$ doublet $\Theta_{\dalpha}$, with $\dalpha = 1, 2$, and are vectors of $\USp{4} \simeq \Spin{5}$. The subscript $R$ is used to emphasise that this $\SU{2}$ corresponds to the $R$-symmetry. Let us begin by fixing our spinor convention.

\subsection{Spinor convention}
The spinors transform as $\USp{4}$ vectors, i.e. we can write $\theta^{\dalpha\,i}$ for each spinor with $i = 1, \ldots, 4$ the $\USp{4}$ index and $\dalpha = 1, 2$ the $\SU{2}_R$ index as discussed above. For $\Spin{5}$ the charge conjugation matrix has to be antisymmetric \cite{VanProeyen:1999ni}. The only invariant tensor we have is the symplectic tensor $\Omega_{ij}$ and so we take this to be the charge conjugation matrix. In particular, it is also unitary because it satisfies
\begin{equation}
 \left(\Omega_{ij}\right)^* = \Omega^{ij} \,.
\end{equation}
Hence
\begin{equation}
 \Omega_{ik} \left(\Omega_{jk}\right)^* = \Omega_{ik} \Omega^{jk} = \delta_i^j \,,
\end{equation}
which shows that it is unitary, i.e. $\Omega \Omega^\dagger = 1$.

Because the charge conjugation matrix is antisymmetric we cannot define Majorana spinors. Instead we can define symplectic Majorana spinors because we have extended SUSY. Thus we have
\begin{equation}
 \left( \theta^* \right)_{\dalpha\,i} = \theta^{\dbeta\,j} \epsilon_{\dbeta\dalpha} \Omega_{ji} \,.
\end{equation}
Thus we will throughout use pseudo-real objects where both the $\USp{4}$ and $\SU{2}_R$ indices are raised/lowered by complex conjugation.

Finally, let us normalise our spinors. We will take the EFT spinors to have weight $-1/10$, matching the usual EFT convention \cite{Godazgar:2014nqa,Musaev:2014lna,Baguet:2016jph}. The full 11-d fermions are a product of the 7-d fermions and the internal spinors and should have no weight. Thus we take the internal spinors $\theta^{\dalpha\,i}$ to have weight $1/10$ and impose the normalisation condition
\begin{equation}
 \theta^{\dalpha\,i} \theta^{\dbeta\,j} \Omega_{ij} = \kappa \epsilon^{\dalpha\dbeta} \,, \label{eq:IntSpinorNormalisation}
\end{equation}
with $\kappa$ a density of weight $1/5$. This looks perhaps more natural if written as a positive definite product:
\begin{equation}
 \theta^{\dalpha\,i} \ctheta_{\dbeta\,i} = \kappa \delta^\dalpha_\dbeta \,.
\end{equation}

\subsection{Spinor bilinears}

We can use these two spinors to construct a set of bilinears which define the $\SU{2}$ structure. In particular, we can form the following pseudo-real $\USp{4}$ tensors
\begin{equation}
 \kappa \,, \qquad A^{ij} = \frac1\kappa \theta^{\dalpha\,i} \theta^{\dbeta\,j} \epsilon_{\dalpha\dbeta} - \frac{1}{2} \Omega^{ij} \,, \qquad B_{u}^{ij} = \frac{i}{\kappa} \theta^{\dalpha\,i} \theta^{\dbeta\,j} \left(\sigma_u\right)_{\dalpha\dbeta} \,, \label{eq:SpinorBilinears}
\end{equation}
where $u = 1, \ldots, 3$ and $\sigma_{u}$ are Pauli matrices. One can check that $A^{ij} \in \mathbf{5}$ and $B_{u}{}^{ij} \in \mathbf{10}$. These tensors satisfy a set of compatibility conditions:
\begin{equation}
 B_{u}{}^{[i}{}_{k} B_{v}{}^{j]k} = A^{ij} \delta_{uv} \,, \qquad B_{u}{}^{[i}{}_k A^{j]k} = 0 \,, \qquad B_{u}{}^{ij} B_{v\,ij} = 2 \delta_{uv} \,. \label{eq:USp4Compatibility}
\end{equation}
Any set of such tensors of $\USp{4}$ have stabiliser $\SU{2} \subset \USp{4}$ and thus define a generalised $\SU{2}$ structure. This can also be seen as follows. Consider decomposing $\USp{4} \longrightarrow \SU{2}_S \times \SU{2}_R$, where $\SU{2}_S$ denotes the $\SU{2}$ structure group. The relevant representations then decompose as
\begin{equation}
 \begin{split}
  \mbf{5} &\longrightarrow \left(\mbf{2},\mbf{2}\right) \oplus \left(\mbf{1},\mbf{1}\right) \,, \\
  \mbf{10} &\longrightarrow \left(\mbf{1},\mbf{3}\right) \oplus \left(\mbf{3},\mbf{1} \right) \oplus \left(\mbf{2},\mbf{2}\right) \,,
 \end{split}
\end{equation}
and $A^{ij}$ and $B_u{}^{ij}$ correspond to the singlets under $\SU{2}_S$.

In order to define a reduction of the $\SL{5} \times \mathbb{R}^+$ structure group to $\SU{2}$ we need to lift these objects to tensors of $\SL{5} \times \mathbb{R}^+$. We then have a $\SL{5}$ vector $A^a$ and three $\SL{5}$ antisymmetric tensors $B_{u,ab}$, which satisfy
\begin{equation}
 B_{u,ab} A^b = 0 \,, \qquad B_{u,ab} B_{v,cd} \epsilon^{abcde} = 4\sqrt{2} \delta_{uv} A^e \,.
\end{equation}
Note that here we use the conventions that
\begin{equation}
 \epsilon^{(ij),(kl)}{}_{[mn]} = 4\sqrt{2} \left( \delta_{[m}{}^{(i} \Omega^{j)(k} \delta_{n]}{}^{l)}  + \frac14 \Omega_{mn} \Omega^{i(k} \Omega^{l)j} \right) \,.
\end{equation}
However, we see that it is impossible to impose all the compatibility conditions \eqref{eq:USp4Compatibility} as $\SL{5}\times\mathbb{R}^+$ relations on $A^a$ and $B_{u,ab}$.

As a result, the objects $\left(\kappa, A^a, B_{u,ab}\right)$ are stabilised by $\SU{2} \ltimes \mathbb{R}^4 \subset \SL{5} \times \mathbb{R}^+$. Indeed, one can check that they define 18 degrees of freedom, which matches the dimension of the coset space
\begin{equation}
 \frac{\SL{5} \times \mathbb{R}^+}{\SU{2} \ltimes \mathbb{R}^4} \,.
\end{equation}

The issue here is that by using spinor bilinears to construct invariant tensors, we are already taking the structure group to be a subgroup of $\USp{4}$. Thus the spinor bilinears can be used to further reduce the structure group from $\USp{4}$ to $\SU{2}$. However, to define a $\SU{2}$ structure of $\SL{5} \times \mathbb{R}^+$, without explicitly requiring the existence of spinors, we need to introduce another $\SL{5}$ covector $A_a$ satisfying
\begin{equation}
 A_a A^a = \frac{1}{2} \,.
\end{equation}
The additional four degrees of freedom of $A_a$ are used to absorb the $T^4$ factor of the stabiliser and thus make the structure group a subgroup of $\USp{4}$. Indeed, the set
\begin{equation}
 \left( \kappa\,, \quad A^a \,, \quad A_a \,, \quad B_{u,ab} \right) \,,
\end{equation}
satisfying
\begin{equation}
 A^a A_a = \frac{1}{2} \,, \qquad B_{u,ab} A^b = 0 \,, \qquad B_{u,ab} B_{v,cd} \epsilon^{abcde} = 4\sqrt{2}\delta_{uv} A^e \,, \label{eq:Compatibility}
\end{equation}
parameterises the coset space
\begin{equation}
 \frac{\SL{5} \times \mathbb{R}^+}{\SU{2}} \,,
\end{equation}
and thus defines a $\SU{2} \subset \SL{5} \times \mathbb{R}^+$ structure. Because $\SU{2} \subset \USp{4}$, this set of tensors also implicitly defines a generalised metric.

To see that the stabiliser is indeed $\SU{2}$, note that up to a $\SL{5} \times \mathbb{R}^+$ transformation we can take
\begin{equation}
 A^5 = \frac{1}{\sqrt{2}} \,, \qquad A^{\bar{i}} = 0 \,, \qquad \bar{i} = 1, \ldots, 4 \,,
\end{equation}
and thus
\begin{equation}
 A_5 = \frac{1}{\sqrt{2}} \,.
\end{equation}
This configuration is stabilised by $\SL{4} \ltimes T^4$ but the $T^4$ degrees of freedom can be used to set $A_{\bar{i}} = 0$. As a result, the stabiliser now becomes $\SL{4} \subset \SL{5} \times \mathbb{R}^+$. The constraint
\begin{equation}
 B_{u,ab} A^b = 0 \,,
\end{equation}
implies that $B_{u,\bar{i}5} = 0$ and thus we are left to satisfy
\begin{equation}
 B_{u,\bar{i}\bar{j}} B_{v,\bar{k}\bar{l}} \epsilon^{\bar{i}\bar{j}\bar{k}\bar{l}} = 4 \delta_{uv} \,.
\end{equation}
Three such antisymmetric rank-two tensors of $\SL{4}$ parameterise the coset space $\SL{4}/\SU{2}$, see e.g. \cite{Triendl:2009ap}.

One should think of the objects $A_a$, $A^a$ and $B_{u,ab}$ as the ``exceptional generalisation'' of the complex and K\"ahler structure on four-manifolds. We have already mentioned that they implicitly define a generalised metric, although there is no explicit formula relating the two. This is not surprising since, for example, the K\"ahler metric on K3 surfaces (which are an example of exceptional $\SU{2}$-manifolds) is not known. However, by their definition we can identify $A_a$ and $B_{u,ab}$ with certain components of the coset representatives $\Vi_a{}^{ij} \in \frac{\SL{5}}{\USp{4}}$. This arises because the coset representatives define the map between $\SL{5}$ and $\USp{4}$ and thus
\begin{equation}
 A_a = \frac1{2\kappa} \Vi_a{}^{ij} \theta_{\dalpha\,i} \theta_j{}^{\dalpha} \,, \qquad B_{u,ab} = \frac{i}{2\kappa} \left(\sigma_u\right)_{\dalpha\dbeta} \Vi_{ab}{}^{ij} \theta_{i}{}^{\dalpha} \theta_j{}^{\dbeta} \,, \label{eq:ABVi}
\end{equation}
where $\Vi_{ab}{}^{ij} = \frac{1}{\sqrt{2}} \Vi_{[a}{}^{i}{}_{|k|} \Vi_{b]}{}^{jk}$.

\subsection{Properties of the spinor bilinears} \label{s:SpinorBilinearProps}
Using $A_a$ we can actually ``raise'' the indices on $B_{u,ab}$ by defining the tensor
\begin{equation}
 V_u{}^{ab} = \epsilon^{abcde} B_{u,cd} A_e \,.
\end{equation}
Due to the compatibility conditions \eqref{eq:Compatibility}, it further satisfies
\begin{equation}
 \frac12 V_u{}^{ab} B^v{}_{ab} = \sqrt{2} \delta_u{}^v \,, \qquad V_u{}^{ab} A_b = 0 \,.
\end{equation}
The generalised vector of weight $\frac15$
\begin{equation}
 \tV_u{}^{ab} = \kappa V_u{}^{ab} \,.
\end{equation}
will play an important role in defining the intrinsic torsion.

Furthermore, using $A^a$ and $A_a$ we can project any $\SL{5}$ vector, $Q^a$, onto a subspace parallel to $A^a$ and perpendicular to it by
\begin{equation}
 Q^a = A^a A_b Q^b + P_a{}^b Q^b \,,
\end{equation}
where we introduced the projector
\begin{equation}
 P_a{}^b = \left( \delta^a{}_b - 2 A^a A_b \right) \,, \qquad P_a{}^b A_b = 0 \,.
\end{equation}
Note that this can also be expressed in terms of $V_u{}^{ab}$ and $B^u{}_{ab}$ as
\begin{equation}
 P_a{}^b = \frac{\sqrt{2}}{3} B^u{}_{ac} V_u{}^{bc} \,.
\end{equation}

One can also form the following objects which are adjoint-valued:
\begin{equation}
 T^u{}_{a}{}^b = \frac1{\sqrt{2}} \epsilon^{uvw} B{}_{v,ac} V_w{}^{bc} \,. \label{eq:HCStructure}
\end{equation}
These satisfy the following algebra
\begin{equation}
 T^u{}_{a}{}^c T^v{}_{c}{}^b = - \delta^{uv} \left( \delta_a{}^b - 2 A_a A^b \right) - \epsilon^{uv}{}_{w} T^w{}_a{}^b \,. \label{eq:HCAlgebra}
\end{equation}
It is clear that these objects form a hyper-complex structure in the subspace perpendicular to $A_a$. In the fluxless M-theory limit this reduces to the hyper-complex structure on 4-manifolds of $\SU{2}$-structure.

Finally, we can also define a metric on the subspace perpendicular to $A_a$ using
\begin{equation}
 \begin{split}
  M_{ab} &= \epsilon_{uvw} B^u{}_{ac} B^v{}_{bd} V^{w,cd} \,, \\
  M^{ab} &= \epsilon^{uvw} V_u{}^{ac} V_v{}^{bd} B_{w,cd} \,, \label{eq:22Metric}
 \end{split}
\end{equation}
which satisfy
\begin{equation}
 M^{ac} M_{cb} = 9 \sqrt{2} P^a{}_b \,.
\end{equation}

\section{Reformulating the $\SL{5}$ EFT}\label{s:Reformulate}
We will now reformulate the $\SL{5}$ EFT in terms of the generalised $\SU{2}$-structure, $\kappa$, $A_a$, $A^a$, $B_{u,ab}$. This can be thought of as an $\SL{5}$ version of the rewriting in \cite{Grana:2005ny,Grana:2006hr,Grana:2009im}, but extended to the case where there are non-vanishing gauge fields. This will be necessary in order to obtain the full gauged SUGRAs after performing a consistent truncation and not just their vacua.

To perform the reformulation we need to introduce a generalised $\SU{2}$-connection, which in general is not torsion-free. To motivate this, consider the case of maximal supersymmetry \cite{Berman:2013uda,Blair:2014zba}. In that instance the consistent truncation is defined on a space with generalised identity-structure \cite{Coimbra:2012af,Lee:2014mla} and thus the compatible connection is uniquely given by the Weitzenb\"ock connection. As showed in \cite{Berman:2013uda,Blair:2014zba} the EFT scalar potential can be rewritten in terms of the torsion of this connection and upon truncation, the torsion becomes the embedding tensor of the maximal gauged SUGRA.

Here we perform the analogous construction in the case of generalised $\SU{2}$-structures for which the connection is not unique. Nonetheless, the intrinsic $\SU{2}$-torsion, which we define and discuss in \ref{s:IntrinsicTorsion}, corresponds to generalised fluxes and can be used to reformulate the theory. For example, the intrinsic torsion appears in the SUSY variations, as we show in \ref{s:SUSYVariations}, and in section \ref{s:ScalarPotential} we show that we can express the scalar potential completely in terms of the intrinsic torsion.\footnote{For readers who wish to read more about intrinsic torsion we refer to \cite{Joyce} as well as \cite{Coimbra:2014uxa} for its uses in generalised geometry.} In section \ref{s:KinTerms} we rewrite the kinetic and topological terms in terms of the generalised $\SU{2}$-structure.

\subsection{Intrinsic $\SU{2}$-torsion} \label{s:IntrinsicTorsion}
We now introduce a $\SU{2}$-connection whose intrinsic torsion will be identified with the embedding tensor of the half-maximal gauged supergravity obtained after truncating. A $\SU{2}$-connection $\tnabla_{ab}$ is compatible with the tensors defining the $\SU{2}$-structure, i.e.
\begin{equation}
 \tnabla_{ab} \kappa = \tnabla_{ab} A^c = \tnabla_{ab} A_c = \tnabla_{ab} B_{u,cd} = 0 \,.
\end{equation}
This does not uniquely specify the connection, unlike in the maximally-supersymmetric case of an identity structure.

Recall from section \ref{s:Overview} that the torsion of a connection $\nabla$ is the tensor part of the connection and can be defined in terms of the generalised Lie derivative, by
\begin{equation}
 \gL_\xi^\nabla U^a - \gL_\xi^\partial U^a = \frac12 \xi^{bc} U^d \tau_{bc,d}{}^a + \frac{\lambda}{2} U^a \xi^{bc} \tau_{bc} \,, \label{eq:LieTorsion2}
\end{equation}
where $U^a$ has weight $\lambda$ under generalised diffeomorphisms and $\tau_{ab}$ is the trombone part of the embedding tensor. We know that for $\SL{5}$ the torsion has components only in the
\begin{equation}
 \tau \in W = \mathbf{15} \oplus \obf{40} \oplus \mbf{10} \,.
\end{equation}

In the following discussion of the intrinsic $\SU{2}$-torsion we essentially follow the general prescription outlined in \cite{Coimbra:2014uxa}. The torsion map viewed as a map from the space of $\SU{2}$ connections to the space of torsions $W$ may be neither injective nor surjective. In the first case, many $\SU{2}$ connections could have the same torsion, while in the latter, it is impossible to find a $\SU{2}$-connection yielding an arbitrary torsion (the torsion map is not right-invertible on all of $W$). The part of the torsion that is independent of the choice of $\SU{2}$ connection is called the \emph{intrinsic} torsion, and is non-zero when the torsion map is not surjective.

\subsubsection{Representations in the intrinsic $\SU{2}$-torsion} \label{s:IntrinsicRepTheory}
To calculate the representations in which the intrinsic $\SU{2}$-torsion transforms note that any two $\SU{2}$ connections must differ by an adjoint valued tensor in the $\obf{10}$, i.e. by $\Sigma \in \Gamma(K_{\SU{2}})$ where
\begin{equation}
 K_{\SU{2}} = \obf{10} \otimes ad(\tilde{P}_{\SU{2}}) \,.
\end{equation}
In terms of $\SU{2}_S \times \SU{2}_R$ representations\footnote{We will be sloppy here and not differentiate between sections and linear representation spaces.} we have
\begin{equation}
 K_{\SU{2}} = \left(\mbf{1},\mbf{1}\right) \oplus \left(\mbf{5},\mbf{1}\right) \oplus \left(\mbf{3},\mbf{1}\right) \oplus \left(\mbf{3},\mbf{3}\right) \oplus \left(\mbf{4},\mbf{2}\right) \oplus \left(\mbf{2},\mbf{2}\right) \,.
\end{equation}
Now the torsion map is a map
\begin{equation}
 \tau: K_{\SU{2}} \longrightarrow W \,,
\end{equation}
where $W = \mbf{15} \oplus \obf{40} \oplus \mbf{10}$ in terms of $\SL{5}$ representations. Decomposing $W$ into $\SU{2}_S \times \SU{2}_R$ we find
\begin{equation}
 W = 2\cdot\left(\mbf{3},\mbf{3}\right) \oplus 2\cdot \left(\mbf{3},\mbf{1}\right) \oplus 2 \cdot \left(\mbf{1},\mbf{3}\right) \oplus 4 \cdot \left(\mbf{2},\mbf{2}\right) \oplus \left(\mbf{2},\mbf{4}\right) \oplus \left(\mbf{4},\mbf{2}\right) \oplus 3 \cdot \left(\mbf{1},\mbf{1}\right) \,.
\end{equation}
Thus we see that the image of the torsion map is at most
\begin{equation}
 W_{\SU{2}} = \textrm{Im} \, \tau \subset \left(\mbf{3},\mbf{1}\right) \oplus \left(\mbf{3},\mbf{3}\right) \oplus \left(\mbf{1},\mbf{1}\right) \oplus \left(\mbf{4},\mbf{2}\right) \oplus \left(\mbf{2},\mbf{2}\right) \,, \label{eq:ImTau}
\end{equation}
and hence the set which is independent of the connection is given by
\begin{equation}
 W_{int} = W / W_{\SU{2}} \supset 2 \cdot \left(\mbf{1},\mbf{1}\right) \oplus \left(\mbf{3},\mbf{1}\right) \oplus 2 \cdot \left(\mbf{1},\mbf{3}\right) \oplus \left(\mbf{3},\mbf{3}\right) \oplus 3 \cdot \left(\mbf{2},\mbf{2}\right) \oplus \left(\mbf{2},\mbf{4}\right) \,. \label{eq:IntrinsicTorsionExpect}
\end{equation}
Finally, for the sake of completeness let us mention that the kernel of the torsion map is at least
\begin{equation}
 U = \textrm{Ker} \, \tau \supset \left(\mbf{5},\mbf{1}\right) \,,
\end{equation}
although this will not concern us further.

In principle the image of $\tau$ could be smaller than the right-hand side of \eqref{eq:ImTau} in which case the intrinsic torsion is larger than the right-hand side \eqref{eq:IntrinsicTorsionExpect}. However, a direct calculation shows that this is not the case and so we find
\begin{equation}
 W_{int} = 2 \cdot \left(\mbf{1},\mbf{1}\right) \oplus \left(\mbf{3},\mbf{1}\right) \oplus 2 \cdot \left(\mbf{1},\mbf{3}\right) \oplus \left(\mbf{3},\mbf{3}\right) \oplus 3 \cdot \left(\mbf{2},\mbf{2}\right) \oplus \left(\mbf{2},\mbf{4}\right) \,. \label{eq:IntrinsicTorsion}
\end{equation}
Because they are intrinsic these are the only components of the $\SU{2}$-torsion that are physically relevant and we will see that these are related to the embedding tensor after truncation. In the following section we will show that the EFT can be rewritten entirely in terms of the $\SU{2}$-structure and its intrinsic torsion. 

\subsubsection{Explicit expressions for the intrinsic $\SU{2}$-torsion}
We now wish to find explicit expressions for the intrinsic $\SU{2}$-torsion which has irreducible components transforming in the representations \eqref{eq:IntrinsicTorsion}. The fact that the intrinsic torsion is independent of the $\SU{2}$-connection means that it can be expressed directly in terms of the $\SU{2}$-structure and its partial derivatives with no connection appearing. Thus, the intrinsic torsion is given by $\SL{5}$ tensorial combinations of derivatives of the $\SU{2}$-structure.

For example, consider the combination
\begin{equation}
 S = A^a \partial_{ab} A^b \,. \label{eq:S1}
\end{equation}
It follows from the tensor hierarchy \cite{Wang:2015hca} that this is a tensor under generalised diffeomorphisms. Regardless we could have written it in terms of any connection $\tnabla$
\begin{equation}
 S = A^a \left( \tnabla_{ab} A^b - \tGamma_{ab,c}{}^b A^c \right) \,,
\end{equation}
where $\tGamma_{ab,c}{}^d$ are the components of the connection $\tnabla$. Because $S$ is a tensor and the first term on the right-hand side of the above equation is a tensor, the final term must be a tensor too. By definition, it is part of the torsion of $\tnabla$. If we now specialise to the case where $\tnabla$ is a $\SU{2}$-connection we find that
\begin{equation}
 S = - A^a \tGamma_{ab,c}{}^b A^c \,,
\end{equation}
where as we said the right-hand side is part of the torsion. However, $S$ was defined in \eqref{eq:S1} without referring to a specific $\SU{2}$-connection and thus we see that it corresponds to the intrinsic torsion.

In order to find expressions for the intrinsic torsion let us first define the projectors onto the representations appearing in \eqref{eq:IntrinsicTorsion}. Firstly, note that $A^a$ define the singlets in the decomposition
\begin{equation}
 \mbf{5} \rightarrow \left(\mbf{1},\mbf{1}\right) \oplus \left(\mbf{2},\mbf{2}\right) \,,
\end{equation}
as $\SL{5} \rightarrow \SU{2}_S \times \SU{2}_R$, and similarly $A_a$ for the $\obf{5}$ decomposition. Then the subspace perpendicular to $A^a$ in the $\mbf{5}$ corresponds to the $\left(\mbf{2},\mbf{2}\right)$. Thus, we use $A_a$ to project onto the $\left(\mbf{1},\mbf{1}\right)$ and the projector we have met in section \ref{s:SpinorBilinearProps} for the $\left(\mbf{2},\mbf{2}\right)$:
\begin{equation}
 \begin{split}
  P_a{}^b &= \delta_a{}^b - 2  A_a A^b \\
  &= \frac{\sqrt{2}}{3} B^u{}_{ac} V_u{}^{bc} \,,
 \end{split}
\end{equation}
and similarly of course for the conjugate $\SL{5}$ reps.

For the $\mbf{10}$ of $\SL{5}$ we have the decomposition
\begin{equation}
 \mbf{10} \rightarrow \left(\mbf{3},\mbf{1}\right) \oplus \left(\mbf{1},\mbf{3}\right) \oplus \left(\mbf{2},\mbf{2}\right) \,,
\end{equation}
as we break $\SL{5} \rightarrow \SU{2}_S \times \SU{2}_R$. The three tensors $V_{u}{}^{ab}$ project onto the $\left(\mbf{1},\mbf{3}\right)$ representations, while $A^a$ can be used to project onto the $\left(\mbf{2},\mbf{2}\right)$. Finally, we can use
\begin{equation}
 P_{ab}{}^{cd} = \left( \delta_{ab}{}^{cd} - \frac1{2\sqrt{2}} B_{u,ab} V_u{}^{cd} + 4 A_{[a} A^{[c} \delta_{b]}{}^{d]} \right) \,,
\end{equation}
to project onto the $\left(\mbf{3},\mbf{1}\right)$ since
\begin{equation}
 P_{ab}{}^{cd} B_{u,cd} = 0 \,.
\end{equation}

Before giving the explicit expressions for the intrinsic torsion let us also define the projector onto the $\left(\mbf{2},\mbf{4}\right) \subset \left(\mbf{2},\mbf{2}\right) \otimes \left(\mbf{1},\mbf{3}\right)$
\begin{equation}
 P_{a}{}^{u,b}{}_{v} = \delta_a{}^b \delta^u{}_v + \frac{\sqrt{2}}{3} B^u{}_{ac} V_v{}^{cb} \,.
\end{equation}

We are now ready to give explicit expressions for the intrinsic torsion.

\paragraph{Singlets}
\begin{equation}
 \begin{split}
  S &=  A^a \partial_{ab} A^b \,, \\
  T &= \frac1{12\kappa} \epsilon_{uvw} V^{u,cd} \gL_{\tV^v} B^w{}_{cd} \,.
 \end{split} \label{eq:TSinglets}
\end{equation}

\paragraph{$\left(\mbf{1},\mbf{3}\right)$}
\begin{equation}
 \begin{split}
  T_u &= - 2 \kappa^2 A^a \gL_{\tV_u} \left(A_a \kappa^{-3} \right) \,, \\
  S_u &= 2 \kappa^{-6} \gL_{\tV_u} \kappa^5 \,.
 \end{split} \label{eq:T13}
\end{equation}

\paragraph{$\left(\mbf{3},\mbf{1}\right)$}
\begin{equation}
 \begin{split}
  T_{ab} &= \frac1{12\kappa} P_{ab}{}^{cd} \gL_{\tV_u} B^u{}_{cd} \\
  &= \frac1{12\kappa} \left( \gL_{\tV_u} B^u{}_{ab} - \frac{1}{2\sqrt{2}} B^v{}_{ab} V_v{}^{cd} \gL_{\tV_u} B^u{}_{cd} + 4 A^c A_{[a} \gL_{\tV_u} B^u{}_{b]c} \right) \,. \label{eq:T31}
 \end{split}
\end{equation}

\paragraph{$\left(\mbf{3},\mbf{3}\right)$}
\begin{equation}
 \begin{split}
  T^u{}_{ab} &= \frac{1}{12\kappa} \epsilon^{uvw} P_{ab}{}^{cd} \gL_{\tV_v} B_{w,cd} \\
  &= \frac1{12\kappa} \epsilon^{uvw} \left( \gL_{\tV_v} B_{w,ab} - \frac1{2\sqrt{2}} B^x{}_{ab} V_x{}^{cd} \gL_{\tV_v} B_{w,cd} + 4 A^c A_{[a} \gL_{\tV_v} B_{|w|,b]c} \right) \,. \label{eq:T33}
 \end{split}
\end{equation}

\paragraph{$\left(\mbf{2},\mbf{2}\right)$}
\begin{equation}
 \begin{split}
  S_a &= \frac{1}{\kappa^3} \partial_{ab} \left( A^b \kappa^3 \right) - 2 A_a A^b \partial_{bc} A^c \,, \\
  T_a &= \frac{1}{12\kappa} \epsilon^{uvw} B_{u,ab} V_v{}^{bc} \gL_{\tV_w} A_c \,, \\
  U_a &= \frac1{\kappa} B_{u,ab} \gL_{\tV^u} A^b \,.
 \end{split} \label{eq:TDoublets}
\end{equation}

\paragraph{$\left(\mbf{2},\mbf{4}\right)$}
\begin{equation}
 T^u{}_a = \frac{1}{\kappa} P_{a}{}^{u,b}{}_{v} \epsilon^{vwx} B_{w,bc} \gL_{\tV_x} A^c \,, \label{eq:TDoublets4}
\end{equation}
or more explicitly
\begin{equation}
 T^u{}_{a} = \frac{1}{\kappa} \left( \epsilon^{uvw} B_{v,ab} \gL_{\tV_w} A^b + \frac{\sqrt{2}}{3} B^u{}_{ab} W^b \right) \,,
\end{equation}
with
\begin{equation}
 W^a = \epsilon^{uvw} V_u{}^{ab} B_{v,bc} \gL_{\tV_w} A^c \,.
\end{equation}

Note that while one can think of other tensorial combinations transforming in the above representations they cannot be linearly independent from the expressions given above. For example, we can of course raise and lower the $\left(\mbf{2},\mbf{2}\right)$ indices using the metric $M_{ab}$ defined in \eqref{eq:22Metric}. We can also dualise the $\left(\mbf{3},\mbf{1}\right)$ and $\left(\mbf{1},\mbf{3}\right)$ indices using $\epsilon^{abcde} A_e$. However, in this case it is clear that the resulting expressions are linearly dependent on the intrinsic torsion given above.

\subsubsection{Intrinsic torsion in terms of spinors}
In order to rewrite the supersymmetry variations it will be useful to express the intrinsic torsion in terms of the spinors $\theta^{\dalpha,i}$. We do this using the torsion-free $\USp{4}$ connection. For example, this allows us to write
\begin{equation}
 \begin{split}
  S &= A^a \partial_{ab} A^b = A^a \nabla_{ab} A^b \\
  &= \frac1{4} \Vi^a{}_{ij} \Vi^{b}{}_{kl} \Vi_{ab}{}^{mn} A^{ij} \nabla_{mn} A^{kl} \\
  &= \frac1{2\sqrt{2}} A^{ij} \nabla_{ik} A^{k}{}_j \
  \,,
 \end{split}
\end{equation}
where $\nabla_{ab}$ is the torsion-free $\USp{4}$ connection as discussed in section \ref{s:Overview}. This can then be expressed in terms of the spinors $\theta^{\dalpha,i}$ by the definition of $A^{ij}$ in equation \eqref{eq:SpinorBilinears}.

One finds that
\begin{equation}
 \begin{split}
  S &= \frac{1}{\sqrt{2}\kappa} \left( \theta^i{}_{\dalpha} \nabla_{ij} \theta^{j,\dalpha} - \frac1{\kappa} \theta^i{}_{\dalpha} \theta^{j,\dbeta} \theta_{k,\dbeta} \nabla_{ij} \theta^{k,\dalpha} \right) \,, \\
  T &= \frac{1}{\kappa} \left( \theta^i{}_{\dalpha} \nabla_{ij} \theta^{j,\dalpha} + \frac1{\kappa} \theta^i{}_{\dalpha} \theta^{j,\dbeta} \theta_{k,\dbeta} \nabla_{ij} \theta^{k,\dalpha} \right) \,, \\
  T_{\dalpha\dbeta} &= i \left(\sigma\right)^{u}_{\dalpha\dbeta} T_u = - \frac{4\sqrt{2}}{\kappa^2} \theta^k{}_{(\dalpha} \theta^i{}_{\dbeta)} \theta^{j,\drho} \nabla_{ij} \theta_{k,\drho} \,, \\
  S_{\dalpha\dbeta} &= i \left(\sigma^u\right)_{\dalpha\dbeta} S_u = \frac{4 \sqrt{2}}{\kappa} \theta^i{}_{(\dalpha} \nabla_{|ij|} \theta^j{}_{\dbeta)} \,.
 \end{split} \label{eq:TorsionSpinors}
\end{equation}

\subsection{Supersymmetry variation of the gravitino}\label{s:SUSYVariations}
Let us begin the rewriting of the theory in terms of the ${\cal N}=2$ structures by studying the supersymmetry variations of the gravitino. The gravitini of the $\SL{5}$ EFT transform in the $\mathbf{4}$ representation of $\USp{4}$. Under $\USp{4} \rightarrow \SU{2}_S \times \SU{2}_R$ this decomposes as
\begin{equation}
 \mathbf{4} \rightarrow \left(\mbf{2},\mbf{1}\right) \oplus \left(\mbf{1},\mbf{2}\right) \,.
\end{equation}
We see that we obtain a $\SU{2}_R$ doublet of gravitini $\psi_\mu{}^{\dalpha}$ as well as $\SU{2}_S$ doublet. The gravitini forming a doublet of $\SU{2}_S$ are responsible for enhancing the SUSY to ${\cal N}=4$ and thus we will ignore them here. Upon imposing the consistent truncation Ansatz they will correspond to massive gravitino multiplets of the gauged SUGRA and we ensure the truncation does not excite them.

By comparison with \cite{Godazgar:2014nqa,Musaev:2014lna} and \cite{Samtleben:2005bp} one can see that the SUSY variation of the gravitini of the $\SL{5}$ EFT can be written (up to coefficients and $\gamma$-matrix orderings which are not important to us here) as
\begin{equation}
 \begin{split}
  \delta_\epsilon \psi^i_\mu &\sim D_\mu \epsilon^i + \Vi^{a\,ik} \Vi^b{}_{jk} \left[ \gamma_\mu \nabla_{ab} \epsilon^j + \gM_{ab} \gM_{cd} \Fa_{\nu\rho}{}^{cd} \gamma^{\nu\rho} \gamma_\mu \epsilon^j \right] \\
  & \quad + \Fb_{\nu\rho\sigma,a} \Vi^a{}_{jk} \Omega^{ij} \gamma^{\nu\rho\sigma} \gamma_\mu \epsilon^k \,.
 \end{split}
\end{equation}

The ${\cal N}=2$ gravitini are embedded in the $\USp{4}$ ones via the internal spinors $\theta^i{}_{\dalpha}$, hence
\begin{equation}
 \tilde{\psi}_\mu{}^i = \theta^i{}_{\dalpha} \psi_\mu{}^{\dalpha} \,.
\end{equation}
The ${\cal N}=2$ SUSY parameters are similarly embedded into the $\USp{4}$ ones as
\begin{equation}
 \tilde{\epsilon}^i = \theta^i{}_{\dalpha} \epsilon^{\dalpha} \,.
\end{equation}

Hence we can write the variation of the ${\cal N}=2$ gravitini as
\begin{equation}
 \begin{split}
  \delta_{\tilde{\epsilon}} \psi_\mu{}^{\dalpha} &\sim - \frac{1}{\kappa}\theta_i{}^{\dalpha} \delta_{\tilde{\epsilon}} \tilde{\psi}_\mu{}^i \\
  &\sim - \frac{1}{\kappa} \left[ \theta_i{}^{\dalpha} D_\mu \left( \theta^i{}_{\dbeta} \epsilon^\dbeta \right) + \theta_i{}^{\dalpha} \Vi^{a\,ik} \Vi^b{}_{jk} \gamma_\mu \nabla_{ab} \left( \theta^j{}_\dbeta \epsilon^{\dbeta} \right) + \theta^j{}_{\dbeta} \theta_i{}^{\dalpha} \Vi_a{}^{ik} \Vi_{b\,jk} \Fa_{\nu\rho}{}^{ab} \gamma^{\nu\rho} \gamma_\mu \epsilon^\dbeta \right. \\
  & \quad \left. - \Fb_{\nu\rho\sigma,a} \theta^{i\,\dalpha} \Vi^a{}_{ij} \theta^j{}_\dbeta \gamma^{\nu\rho\sigma} \gamma_\mu \epsilon^\dbeta \right] \,.
 \end{split}
\end{equation}

In appendix \ref{A:SUSY} we show how one can rewrite this in terms of the $\SU{2}$-structure and its intrinsic torsion. The result is
\begin{equation}
 \begin{split}
  \delta_\epsilon \psi_\mu{}^{\dalpha} &\sim \tilde D_\mu \epsilon^{\dalpha} - \frac1{\kappa} \left(\theta_i{}^{\dalpha} \partial_\mu \theta^i{}_{\dbeta}\right) \epsilon^{\dbeta} + \frac1{20} A_\mu{}^{ab} \tau_{ab} \epsilon^{\dalpha} + \frac1\kappa \theta_i{}^{\dalpha} \left(\gL_{A_\mu}^{\hat{\nabla}} \theta^i{}_{\dbeta}\right) \epsilon^{\dbeta} \\
  & \quad - \frac\kappa2 \left( S + \frac{T}{\sqrt{2}} \right) \gamma_\mu \epsilon^{\dalpha} - \frac{\kappa}{4} S^{\dalpha}{}_{\dbeta} \gamma_\mu \epsilon^{\dbeta} \\
  & \quad - i\sqrt{2} V_u{}^{ab} \left(\sigma^u\right)^{\dalpha}{}_{\dbeta} \gamma_\mu \nabla_{ab} \epsilon^{\dbeta} - i \sqrt{2} B_{u,ab} \Fa_{\nu\rho}{}^{ab} \left(\sigma^u\right)^{\dalpha}{}_{\dbeta} \gamma^{\nu\rho} \gamma_\mu \epsilon^{\dbeta} \\
  & \quad - \Fb_{\nu\rho\sigma,a} A^a \gamma^{\nu\rho\sigma} \gamma_\mu \epsilon^{\dalpha} \,.
 \end{split} \label{eq:N2GravitinoVariation}
\end{equation}

One could proceed similarly for the other fermions which do not form doublets under $\SU{2}_S$ but we will not do so here as this is not necessary for our purposes.

\subsection{Scalar potential}\label{s:ScalarPotential}

It is useful to write the scalar potential as
\begin{equation}
 V = - \frac1{4} \cR - \frac18 \gM^{ac} \gM^{bd} \nabla_{ab} g^{\mu\nu} \nabla_{cd} g_{\mu\nu} \,.
\end{equation}
Here $\cR$ is the so-called generalised Ricci scalar \cite{Hohm:2014qga} -- although it is a density of weight $-\frac{2}{5}$ -- and contains only internal derivatives of the EFT scalars. It can also be written as the square of covariant derivatives of spinors \cite{Coimbra:2011ky,Coimbra:2012af,Hohm:2014qga}, that is
\begin{equation}
 \frac{1}{16} \cR \epsilon^i = \frac{1}{2}\hat{\nabla}_{jk} \hat{\nabla}^{ki} \epsilon^j - \frac12 \hat{\nabla}_{jk} \hat{\nabla}^{jk} \epsilon^i + \frac32 \hat{\nabla}^{ik} \hat{\nabla}_{jk} \epsilon^j \,, \label{eq:VSpinors}
\end{equation}
where $\hat{\nabla}$ is the $\USp{4}$ connection without the seven-dimensional spin connection. This follows from the supersymmetry variation of the fermionic equations of motion which must be proportional to the bosonic equations of motion \cite{Coimbra:2012af}. We show how to derive these coefficients in appendix \ref{A:SpinorPot}.

We now write the spinor as $\epsilon^i = \theta^i{}_{\dalpha} \epsilon^{\dalpha}$ in terms of a $\SU{2}_R$ pair of spinors and use the fact that the right-hand side is linear in $\epsilon^\dalpha$ to find that
\begin{equation}
 \frac{\kappa}{16} \cR = - \theta^{\dalpha}{}_i \left( \frac12 \nabla_{jk} \nabla^{ik} \theta^j{}_{\dalpha} - \frac12 \nabla_{jk} \nabla^{jk} \theta^i{}_{\dalpha} + \frac32 \nabla^{ik} \nabla_{jk} \theta^{j}{}_{\dalpha} \right)\,.
\end{equation}
We further integrate by parts to obtain
\begin{equation}
 \frac{1}{16} \cR = \kappa^{-1} \left( \nabla_{jk} \theta^{\dalpha}{}_i \nabla^{ik} \theta^{j}{}_{\dalpha} - \frac12 \nabla_{jk} \theta^{\dalpha}{}_i \nabla^{jk} \theta^i{}_{\dalpha} + \frac32 \nabla^{ik} \theta^{\dalpha}{}_i \nabla_{jk} \theta^j{}_{\dalpha} \right) \,. \label{eq:PotentialSpinors}
\end{equation}
We will show in \ref{s:SUGRAPotential} that this does reduce to the correct scalar potential of seven-dimensional half-maximal gauged SUGRAs.

Now we are in a position to re-express the potential in terms of the spinor bilinears $A^a$, $A_a$ and $B_{u,ab}$ via their intrinsic torsion \eqref{eq:IntrinsicTorsion}. By expressing the intrinsic torsion in terms of the spinors $\theta^{\dalpha}{}_i$ we find the generalised Ricci scalar to be
\begin{equation}
 \begin{split}
  \cR &= 8\, S^2 - 2\, T^2 - 8\sqrt{2}\, ST - 3\, T_u T^u + T_u S^u - \frac34\, S_u S^u - 16\sqrt{2}\, \epsilon^{abcde} T_{ab} T_{cd} A_e \\
  & \quad - 36\sqrt{2}\, \epsilon^{abcde} T^u{}_{ab} T_{u,cd} A_e - \frac{4\sqrt{2}}{3}\, M^{ab} S_a S_b - \frac{16}{3}\, M^{ab} S_a T_b + \frac83\, M^{ab} U_a S_b \,. \label{eq:ScalarPotential}
 \end{split}
\end{equation}
Here $M^{ab}$ is the metric on the $\left(\mbf{2},\mbf{2}\right)$ as defined in \eqref{eq:22Metric}.

Finally, we claim that one can write
\begin{equation}
 - \frac14 \gM^{ac} \gM^{bd} \nabla_{ab} g_{\mu\nu} \nabla_{cd} g^{\mu\nu} = V_u{}^{ab} V^{u,cd} \tnabla_{ab} g_{\mu\nu} \tnabla_{cd} g^{\mu\nu} \,,
\end{equation}
where $\tnabla_{ab}$ is the $\SU{2}$-connection and which acts on $g_{\mu\nu}$ as
\begin{equation}
 \tnabla_{ab} g_{\mu\nu} = \kappa^2 \partial_{ab} \left( \kappa^{-2} g_{\mu\nu} \right) \,.
\end{equation}
While this term vanishes when performing a consistent truncation as we are doing here, in \cite{Malek:2016vsh} we show that this does reproduce the correct term in the heterotic DFT.

\subsection{Kinetic terms}\label{s:KinTerms}
The kinetic terms of the scalar and gauge fields are usually written in terms of the generalised metric directly
\begin{equation}
 L_{kin} = \frac14 g^{\mu\nu} D_\mu \gM^{ab} D_\nu \gM_{ab} -\frac18 \left( \Fa_{\mu\nu}{}^{ab} \Fa^{\mu\nu,cd} \gM_{ac} \gM_{bd} + \frac{2}{3} \Fb_{\mu\nu\rho,a} \Fb^{\mu\nu\rho}{}_{b} \gM^{ab} \right) \,.
\end{equation}
We need to rewrite these in terms of the $\SU{2}$ structures directly.

It is clear that the kinetic term for the scalars
\begin{equation}
 g^{\mu\nu} D_\mu \gM_{ab} D_\nu \gM^{ab} \,,
\end{equation}
should be replaced by terms involving derivatives of $A^a$, $A_a$ and $B_{u,ab}$. Derivatives of $\kappa$ are of course included in the Einstein-Hilbert term which needs no modification as it does not involve a generalised metric. There are only two such terms which are independent:
\begin{equation}
 g^{\mu\nu} \left( D_\mu B_{u,ab} D_\nu B^u{}_{cd} \right) \epsilon^{abcde} A_e \,, \qquad \textrm{ and } \qquad g^{\mu\nu} D_\mu A^a D_\nu A_a \,.
\end{equation}

Similarly, we wish to replace the kinetic term of the gauge field by the terms
\begin{equation}
 \Fa_{\mu\nu}{}^{ab} \Fa^{\mu\nu,cd} B_{u,ab} B^u{}_{cd} \,, \qquad \textrm{ and } \qquad \Fa_{\mu\nu}{}^{ab} \Fa^{\mu\nu,cd} B_{u[ab} B^u{}_{cd]} \,.
\end{equation}
Note that
\begin{equation}
 B_{u[ab} B^u{}_{cd]} = \frac{1}{\sqrt{2}} \epsilon_{abcde} A^e \,.
\end{equation}
For the $\Fb_{\mu\nu\gamma,a}$ one could consider the term
\begin{equation}
 \Fb_{\mu\nu\gamma,a} \Fb^{\mu\nu\gamma}{}_b A^a A^b
\end{equation}
as well as
\begin{equation}
 \Fb_{\mu\nu\rho,a} \Fb^{\mu\nu\rho}{}_b M^{ab} \,, \label{eq:FbAmbiguity}
\end{equation}
However, as we are about to discuss in the next section \ref{s:Decompostion}, terms such as \eqref{eq:FbAmbiguity} necessarily vanish when we have an honest ${\cal N}=2$ theory and so we will not consider them. This possible omission is irrelevant for ${\cal N}=2$ theories which are the subject of this paper.

We claim that the kinetic terms are given by
\begin{equation}
 \begin{split}
  L_{kin} &= \sqrt{2} g^{\mu\nu} \left( D_\mu B_{u,ab} D_\nu B^{u}{}_{cd} \right) \epsilon^{abcde} A_e - 56\, g^{\mu\nu} D_\mu A^a D_\nu A_a \\
  & \quad + \frac18 \Fa_{\mu\nu}{}^{ab} \Fa^{\mu\nu\,cd} \left( B_{u,ab} B^u{}_{cd} - B_{u[ab} B^u{}_{cd]} \right) - \frac{1}{48} \Fb_{\mu\nu\rho,a} \Fb^{\mu\nu\rho}{}_b A^a A^b \,.
 \end{split} \label{eq:KinTerms}
\end{equation}
One may be able to derive the coefficients appearing here by requiring invariance under external diffeomorphisms. However, we have fixed the coefficients by comparison with gauged SUGRA. As we will see in sections \ref{s:SUGRAKinTerms}, \eqref{eq:KinTerms} does reduce to the correct kinetic terms of seven-dimensional half-maximal gauged SUGRA. In \cite{Malek:2016vsh} we also show that it reproduces the correct kinetic terms of the heterotic DFT.

\section{Consistent truncations to half-maximal gauged supergravity}\label{s:Truncation}

\subsection{Decomposition of supergravity fields}\label{s:Decompostion}
The following discussion is the $\SL{5}$ EFT analogue of the discussion in section 2.2 of \cite{Grana:2005ny} and section 3 of \cite{Grana:2006hr} where they consider four-dimensional ${\cal N}=2$ truncations of 10-dimensional supergravity.

Let us begin by decomposing the EFT fields under $\SU{2}_S \times \SU{2}_R$ where the first factor labels the $\SU{2}$-structure group and the second the $R$-symmetry group. We give the decompositions of the bosons in table \ref{t:BosonDecomp} and that of the fermions in table \ref{t:FermionDecomp}.\\

\noindent\makebox[\textwidth]{
 \begin{minipage}{\textwidth}
  \begin{center}
  \begin{tabular}{|c|c|c|c|}
  \hline
  Field & $\SL{5}$ & $\USp{4}$ & $\SU{2}_S \times \SU{2}_R$ \Tstrut\Bstrut \\ \hline
  $\gM_{ab}$ & $\mbf{15}$ & $\mbf{14} \oplus \mbf{1}$ & $\left(\mbf{3},\mbf{3}\right) \oplus \left(\mbf{1},\mbf{1}\right) \oplus \left(\mbf{2},\mbf{2}\right) \oplus \left(\mbf{1},\mbf{1}\right)$ \\
  $A_\mu{}^{ab}$ & $\mbf{10}$ & $\mbf{10}$ & $\left(\mbf{3},\mbf{1}\right) \oplus \left(\mbf{1},\mbf{3}\right) \oplus \left(\mbf{2},\mbf{2}\right)$ \\
  $B_{\mu\nu\,a}$ & $\obf{5}$ & $\mbf{5}$ & $\left(\mbf{2},\mbf{2}\right) \oplus \left(\mbf{1},\mbf{1}\right)$ \\
  $C_{\mu\nu\rho}{}^a$ & $\mbf{5}$ & $\mbf{5}$ & $\left(\mbf{2},\mbf{2}\right) \oplus \left(\mbf{1},\mbf{1}\right)$ \\
   \hline
  \end{tabular}
 \vskip-0.5em
 \captionof{table}{\small{Decomposition of the $\SL{5}$ EFT bosonic degrees of freedom under $\SU{2}_S \times \SU{2}_R$.}} 
 \label{t:BosonDecomp}
  \end{center}
 \end{minipage}
}
\vspace{1em}

Now we can reorganise all these degrees of freedom into ${\cal N}=2$ supermultiplets. The singlets under the $\SU{2}$-structure group form the graviton supermultiplet.
\begin{equation}
 \textrm{Graviton multiplet:} \quad \left( g_{\mu\nu}, A_{\mu\,\dalpha}{}^\dbeta, \phi, C_{\mu\nu\rho}, \psi_\mu{}^{\dalpha}, \chi^{\dalpha} \right) \,.
\end{equation}
Those in the adjoint of the $\SU{2}$-structure group form the vector multiplets (with $A = 1, \ldots, n$).
\begin{equation}
 \textrm{Vector multiplets:} \quad \left( A_\mu, \phi_\dalpha{}^\dbeta, \chi^{\dalpha} \right)^A \,.
\end{equation}
Finally, all doublets of the $\SU{2}$-structure group form a doublet of gravitino multiplets
\begin{equation}
 \textrm{Gravitino multiplets:} \quad \left( A_\mu{}^{\dalpha}, \phi^{\dalpha}, \psi_\mu, \chi_\dalpha{}^\dbeta, \chi \right)^{\alpha} \,.
\end{equation}

\noindent\makebox[\textwidth]{
 \begin{minipage}{\textwidth}
  \begin{center}
  \begin{tabular}{|c|c|c|}
  \hline
 Field & $\USp{4}$ & $\SU{2}_S \times \SU{2}_R$ \Tstrut\Bstrut \\ \hline
 $\psi_\mu{}^i$ & $\mbf{4}$ & $\left(\mbf{2},\mbf{1}\right) \oplus \left(\mbf{1},\mbf{2}\right)$ \\
 $\chi^{ij,k}$ & $\mbf{16}$ & $\left(\mbf{2},\mbf{1}\right) \oplus \left(\mbf{1},\mbf{2}\right) \oplus \left(\mbf{3},\mbf{2}\right) \oplus \left(\mbf{2},\mbf{3}\right)$ \\
 \hline
  \end{tabular}
 \vskip-0.5em
 \captionof{table}{\small{Decomposition of the $\SL{5}$ EFT fermionic degrees of freedom under $\SU{2}_S \times \SU{2}_R$.}} 
 \label{t:FermionDecomp}
  \end{center}
 \end{minipage}
}
\vspace{1em}

Let us first understand how we obtain $n\neq3$ vector multiplets. A naive expectation would be to have three vector multiplets related to the $\left(\mbf{3},\mbf{1}\right)$ representations. However, the generalised $\SU{2}$-structure group is non-trivially fibred over the manifold. Thus, the number of sections of the $\left(\mbf{3},\mbf{1}\right)$ bundle is in general $n \neq 3$ giving $n\neq3$ vector multiplets. By contrast, the $\SU{2}_R$ group is trivially fibred over the manifold and hence it contains exactly three sections. This is why, for example, we have exactly three vectors in the graviton multiplet and three scalars in each vector multiplet. Finally, the scalars in the vector multiplets, $\phi_{\dalpha}{}^{\dbeta\,A}$, as well as the scalar in the graviton multiplet, $\phi$, will correspond to deformations of the $\SU{2}$ structure $A$ and $B_u$ that we have introduced in section \ref{s:SpinorBilinears}.

Now let us turn to the massive gravitino multiplets. These are associated to broken ${\cal N}=4$ SUSY. Indeed, one can only consistently couple these multiplets to seven-dimensional half-maximal gauged SUGRA for $n=3$ in which case we have a straightforward truncation of a ${\cal N}=4$ theory. Because we want an honest ${\cal N}=2$ theory, we do not want couplings to the gravitino multiplets in the truncated theory. This is ensured by not having any $\SU{2}_S$ doublets in our Ansatz.

One can also understand the need for removing $\SU{2}_S$ doublets in the truncation Ansatz differently. We want to have a generalised $\SU{2}$-structure, not an identity structure. But from the discussion in \ref{s:SpinorBilinears} we see that a nowhere vanishing section in the doublet representation of the $\SU{2}_S$ bundle would correspond to another pair of globally well-defined internal spinors. In this case it is clear that the structure group would be broken to an identity structure and we really have ${\cal N}=4$ SUSY. To avoid this, we project out all doublets of the $\SU{2}$-structure group in our Ansatz.

\subsection{Defining the truncation}\label{s:ExpansionForms}
We now wish to define a consistent truncation of the $\SL{5}$ EFT fields in order to obtain a seven-dimensional half-maximal gauged SUGRA. For the scalar sector we expand the $\SU{2}$-structure $\left(\kappa,\,A_a,\,A^a,\,B_{u,ab}\right)$ in terms of a finite basis of sections which we are about to define.

In the analysis above we have seen that the $\SL{5}$ EFT degrees of freedom organise themselves into sections of the $\left(\mbf{1},\mbf{1}\right)$, $\left(\mbf{1},\mbf{3}\right)$ and $\left(\mbf{3},\mbf{1}\right)$ bundles of $\SU{2}_S \times \SU{2}_R$. Thus we choose a $\SL{5}$ density and a finite number of these sections, which we label by
\begin{equation}
 \rho(Y),\quad n^a(Y),\quad n_a(Y),\quad \omega_{M,ab}(Y) \,,
\end{equation}
and where we have made it explicit that these objects only depend on the internal manifold. $n^a$ and $n_a$ form a basis for the $\left(\mbf{1},\mbf{1}\right)$ sections coming from the $\mbf{5}$ and $\obf{5}$ of $\SL{5}$ respectively. Similarly, the $\omega_{M,ab}$ provide a basis for the $\left(\mbf{3},\mbf{1}\right) \oplus \left(\mbf{1},\mbf{3}\right)$ sections and thus satisfy
\begin{equation}
 \omega_{M,ab} n^b = 0 \,. \label{eq:Constraint1}
\end{equation}
Furthermore they consist of three sections of the $\left(\mbf{1},\mbf{3}\right)$-bundle and $n$ sections of the $\left(\mbf{3},\mbf{1}\right)$-bundle, reflecting the fact that the $\SU{2}_S$ is non-trivially fibred while the $\SU{2}_R$ is trivially fibred, as already discussed in \ref{s:Decompostion}. We can thus write
\begin{equation}
 \omega_{M,ab} = \left( \omega_{I,ab}, \, \omega_{A,ab} \right) \,,
\end{equation}
where $I = 1, 2, 3$ labels the $\SU{2}_R$ adjoint sections and $A = 1, \ldots, n$ labels the $\SU{2}_S$ adjoint sections.

We normalise these sections according to
\begin{equation}
 n^a n_a = 1 \,, \qquad \omega_{M,ab} \omega_{N,cd} \epsilon^{abcde} = 4 \eta_{MN} n^e \,, \label{eq:SectionNormalisation}
\end{equation}
where $\eta_{MN}$ has signature $\left(3,n\right)$ reflecting the number of $adj(\SU{2}_R)$ and $adj(\SU{2}_S)$ sections. We will throughout this paper use $\eta_{MN}$ to raise and lower $\mbf{\left(n+3\right)}$ vector indices. We can use these relations to introduce $n+3$ sections of the $\left(\mbf{3},\mbf{1}\right) \oplus \left(\mbf{1},\mbf{3}\right) \subset \obf{10}$ of $\SL{5}$. These are given by
\begin{equation}
 \omega_M{}^{ab} = \epsilon^{abcde} \omega_{M,cd} n_e \,. \label{eq:Dual10}
\end{equation}
These satisfy
\begin{equation}
 \omega_M{}^{ab} \omega_{N,ab} = 4 \eta_{MN} \,, \qquad \omega_M{}^{ab} n_a = 0 \,. \label{eq:DualNormalisation}
\end{equation}

Given these relationships we can further deduce the following identities which we will use copiously in this paper.
\begin{equation}
 \begin{split}
  \omega_{(M}{}^{cb} \omega_{M)ca} &= \eta_{MN} \left( \delta_a{}^b - n_a n^b \right) \,, \\
  \omega_{(M}{}^{ac} \omega_{P)ab} \omega_N{}^{bd} &= \omega_N{}^{cd} \eta_{MP} \,, \\
  \omega_{M,ab} \epsilon^{abcde} &= 3 \omega_M{}^{[cd} n^{e]} \,, \\
  \omega_M{}^{ab} \epsilon_{abcde} &= 12 \omega_{M[cd} n_{e]} \,, \\
  \omega_M{}^{cd} n^e \epsilon_{abcde} &= 4 \omega_{M,ab} \,, \\
  \omega_M{}^{ab} \omega_N{}^{cd} \epsilon_{abcde} &= 16 \eta_{MN} n^e \,.
 \end{split}\label{eq:OmegaIdentities}
\end{equation}
Furthermore, we will often find it convenient to use the following tensor densities of weight $\frac15$ under generalised diffeomorphisms
\begin{equation}
 \tomega_M{}^{ab} = \rho\, \omega_M{}^{ab} \,.
\end{equation}
In particular, $\tomega_M{}^{ab}$ is a generalised vector and will be useful in formulating the consistency condition for our Ansatz.

Before giving the truncation Ansatz, let us point out that in general we are not developing an effective theory because our truncation Ansatz may be keeping heavy modes, while discarding lower ones.\footnote{This is not because we are using exceptional field theory and thus keeping ``wrapping modes'' but a generic and desired feature of consistent truncation Ans\"atze. Indeed the truncation considered here could equally have been performed in generalised geometry.} Instead we wish to perform a consistent truncation such that all solutions to the equations of motions of the  lower-dimensional theory are also solutions to the equations of motions of the full exceptional field theory, and thus of 11-dimensional supergravity or type IIB. This allows us for example to perform a consistent truncation on a background that is not a solution of the equations of motion.

Because we are only requiring a consistent truncation, not an effective one, the basis of sections which we use for the truncation are not in general analogues of harmonic forms. Indeed, they should not correspond to topological invariants of the background manifold on which we define the consistent truncation. This is because one manifold may admit several different consistent truncations for which different modes are kept, see for example the discussion in the case of maximal SUSY in \cite{Lee:2014mla}. Instead, we will require a weaker set of differential constraints on the sections which we discuss in subsection \ref{s:ConsistencyConditions}.

\subsection{Truncation Ansatz}\label{s:TruncationAnsatz}

\subsubsection{Scalar truncation Ansatz}\label{s:ScalarTruncation}
We begin by expanding the generalised $\SU{2}$ structure in terms of the basis of sections defining the truncation. We let the coefficients in the expansion depend on $x^\mu$, the seven coordinates of the external space. These coefficients determine how the generalised $\SU{2}$-structure, hence the geometry of the internal manifold, changes and they become scalars of the truncated seven-dimensional theory.

We will denote the truncation Ansatz by angled brackets: $\langle \, \rangle$. For the scalar fields it is given by
\begin{equation}
 \begin{split}
  \langle\kappa\rangle(x,Y) &= |\bae|^{1/7}(x)\,e^{-2d(x)/5}\, \rho(Y) \,, \\
  \langle A^a\rangle(x,Y) &= \frac{1}{\sqrt{2}} e^{-4d(x)/5} n^a(Y) \,, \\
  \langle A_a\rangle(x,Y) &= \frac{1}{\sqrt{2}} e^{4d(x)/5} n_a(Y) \,, \\
  \langle B_{u,ab} \rangle(x,Y) &= e^{-2d(x)/5}\, b_{u,M}(x) \omega^M{}_{ab}(Y) \,.
\label{eq:ScalarAnsatz}
 \end{split}
\end{equation}
This implies that
\begin{equation}
 \langle V_u{}^{ab} \rangle = \frac{1}{\sqrt{2}} e^{2d(x)/5}\, b_{u,M}(x)\, \omega^{M,ab}(Y) \,.
\end{equation}

We must now check the compatibility conditions \eqref{eq:Compatibility}. The Ansatz \eqref{eq:ScalarAnsatz} automatically satisfies
\begin{equation}
 A_a A^a = \frac{1}{2} \,.
\end{equation}
However in order to satisfy
\begin{equation}
 B_{u,ab} B_{v,cd} \epsilon^{abcde} = 4 \sqrt{2} A^e \,,
\end{equation}
we find using \eqref{eq:SectionNormalisation} that
\begin{equation}
 b_{u,M} b_{v,N} \eta^{MN} = \delta_{uv} \,. \label{eq:ScalarConstraint}
\end{equation}
This imposes six constraints on the $3n+9$ scalars $b_{u,M}$. Furthermore, it is clear that a rotation on the $u$ index of $b_{u,M}$ corresponds to a $\SU{2}_R$ rotation of the theory. We thus identify any three sets of $b_{u,M}$ related by the action of $\SU{2}_R$. This removes another three degrees of freedom of $b_{u,M}$.

We are left with $3n$ degrees of freedom which is the dimension of the coset space
\begin{equation}
 {\cal M}_{coset} = \frac{\ON{3,n}}{\ON{3}\times\ON{n}} \,.
\end{equation}
Indeed, we can write
\begin{equation}
 b_{u,M} b^{u}{}_{N} = \frac12 \left( \eta_{MN} - \gH_{MN} \right) \,, \label{eq:bH}
\end{equation}
where $\gH_{MN}$ satisfies
\begin{equation}
 \gH_{MP} \gH_{NQ} \eta^{PQ} = \eta_{MN} \,,
\end{equation}
because of \eqref{eq:ScalarConstraint}. Thus $\gH_{MN}$ is a symmetric element of $\ON{3,n}$ and hence gives coordinates on the coset space ${\cal M}_{coset}$. It is the generalised metric of the seven-dimensional gauged supergravity.

There are two further scalars $d(x)$ and $|\bae|(x)$. These are related to the dilaton and the determinant of the seven-dimensional metric $\bag_7$ with $|\bae| = |\bag_7|^{1/2}$. In total we see that we obtain the scalar coset space
\begin{equation}
 {\cal M}_{scalar} = \frac{\ON{3,n}}{\ON{3}\times\ON{n}} \times \mathbb{R}^+ \,,
\end{equation}
where we are not counting $|\bae|$ as part of the scalar manifold because it forms part of the external metric.

\subsubsection{Fermion, gauge field and external metric truncation Ansatz} \label{s:FermionGaugeTruncation}
Let us now give the truncation Ans\"atze for the fermions and gauge fields. Recall from the discussion in \ref{s:SUSYVariations} that the ${\cal N}=2$ gravitini are embedded as $\USp{4}$ fermions by
\begin{equation}
 \begin{split}
  \psi_\mu{}^i &= \theta^i{}_{\dalpha} \psi_\mu{}^{\dalpha} \,. \label{eq:GravitinoEmbedding}
 \end{split}
\end{equation}
Furthermore, we have rewritten the SUSY variations in terms of $\psi_\mu{}^{\dalpha}$, $\chi^{\dalpha}$, $\chi^{\alpha,ij}$ and $\theta_i{}^{\dalpha}$. The truncation Ansatz for these objects is analogous to \eqref{eq:ScalarAnsatz}, e.g. for the gravitino it takes the form
\begin{equation}
 \begin{split}
  \langle \psi_\mu{}^{i} \rangle(x,Y) &= \psi_\mu{}^{\dalpha}(x) \, \Phi^i{}_{\dalpha}(Y)\, \rho^{1/2}(Y) \,, \label{eq:SpinorAnsatz}
 \end{split}
\end{equation}
where $\Phi^i{}_{\dalpha}(Y)$ is now an internal spinor with no weight under the generalised Lie derivative and is the fermionic analogoue of $\omega^M{}_{ab}$, $n^a$ and $n_a$, which can in turn be written as bilinears of $\Phi^i{}_{\dalpha}$. Writing \eqref{eq:SpinorAnsatz} in terms of the ${\cal N}=2$ gravitini and $\theta^i{}_{\dalpha}$ directly it becomes
\begin{equation}
 \begin{split}
  \langle \psi_\mu{}^{\dalpha}\rangle(x,Y) &= \psi_\mu{}^{\dalpha}(x) \,, \\
  \langle \theta^{i}_{\dalpha} \rangle (x,Y) &= \Phi^i{}_{\dalpha}(Y)\, \rho^{1/2}(Y) \,. \label{eq:SpinorAnsatz2}
 \end{split}
\end{equation}

The $\SL{5}$ EFT has one-form, two-form and three-form gauge fields, as well as an auxiliary four-form valued in the $\mbf{10}$, $\obf{5}$, $\mbf{5}$ and $\obf{10}$ of $\SL{5}$ and of weight $\frac15$, $\frac25$, $\frac35$ and $\frac45$ respectively. This determines their truncation Ans\"atze to be
\begin{equation}
 \begin{split}
  \langle \TA_\mu{}^{ab} \rangle (x,Y) &= A_\mu{}^M(x)\, \omega_M{}^{ab}(Y) \, \rho(Y) \,, \\
  \langle \TB_{\mu\nu,a} \rangle (x,Y) &= -4 B_{\mu\nu}(x) \, n_a(Y) \, \rho^2(Y) \,, \\
  \langle \TC_{\mu\nu\gamma}{}^a \rangle (x,Y) &= C_{\mu\nu\gamma}(x) \, n^a(Y) \, \rho^3(Y) \,, \\
  \langle \TD_{\mu\nu\gamma\sigma\,ab} \rangle(x,Y) &= D_{\mu\nu\gamma\sigma\,M}(x)\, \omega^{M}{}_{ab}(Y)\, \rho^4(Y) \,.
 \end{split} \label{eq:GaugeAnsatz}
\end{equation}
The factor of $-4$ in the two-form Ansatz has been chosen to match the half-maximal gauged SUGRA conventions.

Similarly, the truncation Ansatz for the external metric is given by
\begin{equation}
 \langle e_\mu{}^{\bmu} \rangle (x,Y) = \bae_\mu{}^{\bmu}(x)\, e^{-2d(x)/5}\, \rho(Y) \,.
\end{equation}
We have included the power of the dilaton in order to recover the string-frame action.

\subsection{Consistency conditions and the embedding tensor}\label{s:ConsistencyConditions}
We have already listed a set of algebraic constraints which the truncation basis needs to satisfy. These are given by equations \eqref{eq:Constraint1} and \eqref{eq:SectionNormalisation}. However, this is not enough to guarantee a consistent truncation. As we already mentioned, we are in general not truncating to the massless or lowest-lying excitations of a background. Thus, our sections are not some sort of ``exceptional harmonic forms''. Instead we require them to satisfy a weaker set of constraints which can be naturally formulated in terms of the generalised Lie derivative and ensures that we have a consistent truncation.

\subsubsection{Doublet and closure conditions}\label{s:DoubletClosureConditions}
First of all, we must ensure that we do not excite any doublets of $\SU{2}_S$, as we already discussed in \ref{s:Decompostion}. Thus, we require that any doublets generated by the tensorial combinations of derivatives vanish. In particular, we impose
\begin{equation}
 \begin{split}
  n^a \gL_{\tomega_M} \tomega_{N,ab} &= 0 \,, \\
  \gL_{\tomega_M} n^a &= n^a n_b \gL_{\tomega_M} n^b \,, \\
  \partial_{ab} \left( n^b \rho^3 \right) &= \rho^3 n_a n^b \partial_{bc} n^c \,.
 \end{split} \label{eq:DoubletConditions}
\end{equation}
The first equation is manifestly a tensor while it can be checked from \cite{Wang:2015hca} that the second equation is also a tensor. It is easy to see using \eqref{eq:ScalarAnsatz} that these conditions ensure that the doublets of the intrinsic torsion \eqref{eq:TDoublets} and \eqref{eq:TDoublets4} vanish.

Furthermore, we require the sections $\omega_{M,ab}$ to form a closed set under the generalised Lie derivative, i.e.
\begin{equation}
 \gL_{\tomega_M} \omega^N{}_{ab} = \frac14 \left( \gL_{\tomega_M} \omega^N{}_{cd} \right) \omega_P{}^{cd} \omega^P{}_{ab} \,. \label{eq:ClosureCondition}
\end{equation}
In other words, the generalised Lie derivative of $\omega_{M,ab}$ can be expanded in the basis of $\omega_{M,ab}$'s. Using \eqref{eq:DoubletConditions} one can see that this implies
\begin{equation}
 \gL_{\tomega_M} \omega^N{}^{ab} = \frac14 \left( \gL_{\tomega_M} \omega^N{}^{cd} \right) \omega_P{}_{cd} \omega^P{}^{ab} \,,
\end{equation}
so that the $\omega_M{}^{ab}$'s also form a closed set under the generalised Lie derivative.

These conditions are analogous to the differential conditions encountered when studying consistent truncations of $\SU{3}$-structure manifolds \cite{Grana:2005ny,Grana:2006hr}. There one requires the sections used in the truncation Ansatz to form a closed set under the exterior derivative. In the case of consistent maximally supersymmetric truncations of EFT, which are governed by generalised identity-structures, these conditions are satisfied automatically and thus do not need to be imposed by hand.

However, these conditions are not yet enough to guarantee a consistent truncations. The remaining consistency condition is best understood by using the terminology of the embedding tensor to which we turn next.

\subsubsection{The half-maximal embedding tensor}\label{s:EmbeddingTensor}

It is easy to show that the conditions \eqref{eq:SectionNormalisation}, \eqref{eq:DualNormalisation} imply the following identities
\begin{equation}
 \begin{split}
  \gL_{\tomega_M} \omega_N{}^{ab} \omega_{P,ab} &= - \gL_{\tomega_M} \omega_{P,ab} \omega_N{}^{ab} \,, \\
  \gL_{\tomega_M} \omega_N{}^{ab} \omega_{P,ab} - \gL_{\tomega_M} \omega_{N,ab} \omega_P{}^{ab} &= 4 n^a \gL_{\tomega_M} n_a \eta_{NP} \,, \\
  \gL_{\tomega_{M}} \omega_{(N}{}^{ab} \omega_{P),ab} &= 2 n^a \gL_{\tomega_M} n_a \eta_{NP} \,,
 \end{split} \label{eq:LieIdentities}
\end{equation}
where the third equation follows from the first two.

We will now show that the object
\begin{equation}
 g_{MNP} \equiv \frac14 \gL_{\tomega_M} \omega_{N,ab} \omega_{P}{}^{ab} \,, \label{eq:TotalFluxes}
\end{equation}
contains only the irreducible representations allowed by the linear constraint of half-maximal gauged SUGRA \cite{Dibitetto:2012rk} and can thus be identified with the embedding tensor. Let us first define the $\ON{n+3}$ vectors
\begin{equation}
 f_M = n^a \gL_{\tomega_M} n_a \,, \qquad \xi_M = \rho^{-1} \gL_{\tomega_M} \rho \,. \label{eq:VectorFluxes}
\end{equation}

It follows immediately from \eqref{eq:LieIdentities} that
\begin{equation}
 \begin{split}
  g_{M(NP)} = \frac14 \gL_{\tomega_M} \omega_{(N|ab|} \omega_{P)}{}^{ab} &= -\frac12 \eta_{NP}\, n^a  \gL_{\tomega_M} n_a = -\frac12 f_M \eta_{NP} \,,
 \end{split}
\end{equation}
This implies that $g_{MNP} \in \mbf{\left(n+3\right)} \times \left( \mbf{adj} + \mbf{\left(n+3\right)}\right)$. Furthermore, one can use the fact that the torsion lies in the $\mbf{15} \oplus \obf{40} \oplus \mbf{10}$ of $\SL{5}$ to show that
\begin{equation}
 \begin{split}
  g_{(MN)P} &= \frac14 \gL_{\tomega_{(M}} \omega_{N)ab} \omega_{P}{}^{ab} \\
  &= 2 \xi_P \eta_{MN} - 2 \xi_{(M} \eta_{N)P} + \frac14 \eta_{MN} f_P - f_{(M} \eta_{N)P} \,.
 \end{split}
\end{equation}

Thus we see that the only irreducible representations of $g_{MNP}$ are given by
\begin{equation}
 f_{MNP} = g_{[MNP]} \,, \qquad f_M \,, \qquad \xi_M \,. \label{eq:EmbeddingTensor}
\end{equation}
These are exactly the representations allowed for the embedding tensor by the linear constraint of the half-maximal gauged supergravity and we will see that indeed these objects $f_{MNP}$ are to be identified with the embedding tensor. Additionally, there is a singlet deformation allowed in seven-dimensional half-maximal gauged supergravity \cite{Bergshoeff:2007vb} which we identify with
\begin{equation}
 \Theta = \rho\, n^a \partial_{ab} n^b \,. \label{eq:SingletFlux}
\end{equation}
By comparison with \eqref{eq:TSinglets} - \eqref{eq:T33}, $f_{MNP}$, $f_M$, $\xi_M$ and $\Theta$ can also be identified with the intrinsic $\SU{2}$ torsion of the background on which the truncation is defined.

The embedding tensor of gauged SUGRAs has to also satisfy a quadratic constraint which ensures closure of the gauge group. Similarly, consistency of the EFT requires closure of the algebra of generalised Lie derivatives. Indeed by the definition of the embedding tensor in terms of generalised Lie derivatives \eqref{eq:EmbeddingTensor}, \eqref{eq:VectorFluxes}, the closure of the algebra of generalised Lie derivatives automatically implies that the quadratic constraint for the embedding tensor is satisfied.

For example, we could derive a set of quadratic constraints by considering
\begin{equation}
 \left[ \gL_{\tomega_M} ,\, \gL_{\tomega_N} \right] \omega_{P,ab} = \gL_{[\tomega_M,\tomega_N]} \omega_{P,ab} \,,
\end{equation}
where $\left[ \tomega_M, \tomega_N \right]^{ab} \equiv \gL_{\tomega_M} \tomega_N{}^{ab}$. Another set of quadratic constraints comes from
\begin{equation}
 n^a \left[ \gL_{\tomega_M}, \, \gL_{\tomega_N} \right] n_a = n^a \gL_{[\tomega_M,\tomega_N]} n_a \,.
\end{equation}
If we contract with $\eta^{MN}$ the left-hand side vanishes identically whereas the right-hand side gives
\begin{equation}
 \eta^{MN} f_M f_N = 0 \,,
\end{equation}
which indeed reproduces a quadratic constraint for the vector fluxes of half-maximal gauged SUGRA, see e.g. \cite{Dibitetto:2015bia} for the case where $n=3$.

As we already discussed in \ref{s:Overview}, the algebra of generalised Lie derivatives closes when the section condition is fulfilled. Thus, when the background satisfies the section condition, the gaugings automatically satisfy the quadratic constraint. There may however, be examples where the section condition is violated but the quadratic constraint is not.

Furthermore, exactly as in the maximal case \cite{Hohm:2014qga}, we require $\xi_M = 0$ in order to have an action principle for the reduced theory. This can be seen, exactly as in \cite{Hohm:2014qga} by requiring integration by parts to be valid. We want boundary terms to vanish
\begin{equation}
 \int \partial_{ab} ( |e| V^{ab} ) = 0 \,,
\end{equation}
where $|e|$ is the determinant of the external vielbein and $V^{ab}$ has weight $-1/5$ under generalised Lie derivatives. However, we can write
\begin{equation}
 \partial_{ab} \left( |e| V^{ab} \right) = 2 \gL_{\hat{V}} |e|^{5/7} \,,
\end{equation}
where $\hat{V}^{ab} = |e|^{2/7} V^{ab}$ is a generalised vector of weight $\frac15$. After imposing the truncation Ansatz we find
\begin{equation}
 \langle \partial_{ab} \left(|e| V^{ab} \right) \rangle= 10 |e| V^M \xi_M \,,
\end{equation} 
and hence we find that integration by parts is only possible when $\xi_M = 0$. $\xi_M$ is known as the trombone gauging and it is also known from the gauged SUGRA perspective that such a gauging prohibits an action principle \cite{LeDiffon:2008sh}.

By performing the truncation on the SUSY variation in section \ref{s:SUGRASUSY} and particularly scalar potential in section \ref{s:SUGRAPotential}, we will obtain further evidence that the objects $f_{MNP}$, $f_M$, $\xi_M$ and $\Theta$ are to be identified with the embedding tensor. Furthermore, we will see that upon using the reduction Ansatz, all the $\omega^M{}_{ab}$'s, $n^a$'s and $\hat{n}_a$'s will drop out and the only possible dependence on $Y^{ab}$ in the action will appear through the embedding tensor components $f_{MNP}$, $f_M$, $\xi_M$ and $\Theta$ and an overall factor given by a power of the internal density $\rho(Y)$. Thus when the embedding tensor components are constant and obey the quadratic constraint, e.g. by requiring the section condition the internal space, we obtain a consistent truncation to a seven-dimensional gauged SUGRA.

\subsection{Intrinsic torsion and the $T$-tensor}\label{s:SUGRATTensor}
Let us now evaluate the intrinsic torsion \eqref{eq:TSinglets} - \eqref{eq:TDoublets4} using the truncation Ansatz \eqref{eq:ScalarAnsatz}, the relations \eqref{eq:SectionNormalisation}, \eqref{eq:DualNormalisation}, \eqref{eq:DoubletConditions} and the definitions \eqref{eq:EmbeddingTensor}, \eqref{eq:VectorFluxes} and \eqref{eq:SingletFlux}. We immediately find that the doublets \eqref{eq:TDoublets} and \eqref{eq:TDoublets4} vanish on account of \eqref{eq:DoubletConditions}. For the other representations we obtain
\begin{equation}
 \begin{split}
  \langle S \rangle &= \frac{1}{2\rho} e^{-8d/5} \Theta \,, \\
  \langle T \rangle &= \frac{1}{6\rho} e^{2d/5} \gH^{MNP} f_{MNP} \,, \\
  \langle T_u \rangle &= \frac{1}{\rho\sqrt{2}} e^{2d/5} b_u{}^M \left( 3 \xi_M - f_M \right) \,, \\
  \langle S_u \rangle &= \frac{5\sqrt{2}}{\rho} e^{2d/5} b_u{}^M \xi_M \,, \\
  \langle T_{ab} \rangle &= \frac1{8\rho\sqrt{2}}\, P_+^N{}_M \omega^M{}_{ab} \left( 4 \xi_N + f_N \right) \,, \\
  \langle T^u{}_{ab} \rangle &= \frac{1}{12\rho\sqrt{2}}\, \epsilon^{uvw} b_v{}^N b_w{}^P P_+^Q{}_M \omega^M{}_{ab} f_{NPQ} \,.
 \end{split} \label{eq:TTensor}
\end{equation}
Here we defined the left-moving and right-moving projectors $P_-^{MN}$ and $P_+^{MN}$ as well as the antisymmetric tensor $\gH^{MNP}$ as
\begin{equation}
 \begin{split}
  P_-^{MN} &= b_u{}^M b^{u,N} = \frac12 \left( \eta^{MN} - \gH^{MN} \right) \,, \\
  P_+^{MN} &= \eta^{MN} - b_u{}^M b^{u,N} = \frac12 \left( \eta^{MN} + \gH^{MN} \right) \,, \\
  \gH^{MNP} &= \epsilon^{uvw} b_u{}^M b_v{}^N b_w{}^P \,.
 \end{split}
\end{equation}

Similar to the maximally supersymmetric case \cite{Hohm:2014qga} we expect these expressions to correspond to the T-tensor of the seven-dimensional half-maximal gauged supergravity, some components of which are given in \cite{Dibitetto:2015bia} for the case of three vector multiplets, i.e. $n=3$.

\subsection{Reducing the external covariant derivative}\label{s:SUGRAD}
As a first check that we are obtaining a half-maximal gauged SUGRA let us consider the reduction of the external covariant derivative $D_\mu$. We can consider acting with it on any generalised vector, i.e. an object in the $\mbf{10}$ of weight $\frac15$, call it $W^{ab}$ with truncation Ansatz
\begin{equation}
 \langle W^{ab} \rangle (x,Y) = W^M(x) \, \omega_M{}^{ab}(Y) \, \rho(Y) \,.
\end{equation}
Then from equations \eqref{eq:EmbeddingTensor} and \eqref{eq:VectorFluxes} we find
\begin{equation}
 \begin{split}
  \langle D_\mu W^{ab} \rangle (x,Y) &= \rho\, \omega_M{}^{ab} \left( \partial_\mu W^M - \frac1{4\rho} W^N \gL_{A_\mu} \tomega_N{}^{cd} \omega_{M,cd} \right) \\
  &= \rho\, \omega_M{}^{ab} \left( \partial_\mu W^M + A_\mu{}^N W^P g_{NP}{}^M - A\mu{}^N \xi_N W^M \right) \\
  &= \rho\, \omega_M{}^{ab} \D_\mu W^M \,,
 \end{split}
\end{equation}
where $\D_\mu$ is the gauge-covariant derivative of the half-maximal gauged SUGRA (usually this is only given in the case of vanishing 1-form fluxes $f_M = \xi_M = 0$).

Similarly, if we have the external covariant derivative $D_\mu$ acting on a $\SL{5}$ vector $X^a$ which is truncated as
\begin{equation}
 \langle X^a \rangle (x,Y) = X(x)\, n^a (Y) \,,
\end{equation}
then the external covariant derivative reduces as
\begin{equation}
 \langle D_\mu X^a \rangle(x,Y) = n^a \left( \partial_\mu X^a + A_\mu{}^N f_N X \right) = n^a \D_\mu X \,.
\end{equation}
Again $\D_\mu$ corresponds to the gauge-covariant derivative of the gauged supergravity.

\subsection{Reducing the scalar potential}\label{s:SUGRAPotential}
We will now take $\xi_M = 0$ in order to have an action for the reduced theory. Recall that the potential is given by
\begin{equation}
 V = - \frac14 \cR + V_u{}^{ab} V^{u,cd} \nabla_{ab} g_{\mu\nu} \nabla_{cd} g^{\mu\nu} \,.
\end{equation}
It is easy to see that
\begin{equation}
 \langle \nabla_{ab} g_{\mu\nu} \rangle = 0 \,,
\end{equation}
so we are left to evaluate the generalised Ricci scalar.

From equations \eqref{eq:ScalarPotential} and \eqref{eq:TTensor} we can see that the potential of the truncated theory becomes
\begin{equation}
 \begin{split}
  \langle |e| V \rangle &= - \frac14 \rho^5 |\bae| e^{-2d} \left[ P_-^{MQ} P_-^{NR} \left( P_+^{PS} + \frac13 P_-^{PS} \right) f_{MNP} f_{QRS} + \frac12 \left( P_{+}^{MN} + 3 P_{-}^{MN} \right) f_M f_N \right] \\
  & \quad + \frac12 \rho^5 |\bae| e^{-6d} \Theta^2 - \frac{\sqrt{2}}{6} \rho^5 |\bae|e^{-4d} \Theta \gH^{MNP} f_{MNP} \,.
 \end{split}
\end{equation}
By writing out the left-moving and right-moving projectors explicitly and recalling that the quadratic constraint implies $\eta^{MN} f_M f_N = 0$, we obtain
\begin{equation}
 \begin{split}
  \langle |e| V \rangle &= \frac14 \rho^5 |\bae|e^{-2d} f_{MNP} f_{QRS} \left( -\frac1{12} \gH^{MQ} \gH^{NR} \gH^{PS} + \frac14 \eta^{MQ} \eta^{NR} \gH^{PS} - \frac16 \eta^{MQ} \eta^{NR} \eta^{PS} \right) \\
  & \quad - \frac{1}{8} \rho^5 |\bae|e^{-2d} \gH^{MN} f_M f_N + \frac12 \rho^5 |\bae| e^{-6d} \Theta^2 - \frac{\sqrt{2}}{6} \rho^5 |\bae|e^{-4d} \Theta \gH^{MNP} f_{MNP} \,.
 \end{split}
\end{equation}
This is precisely the scalar potential of seven-dimensional half-maximal gauged SUGRA coupled to $n$ vector multiplets, with general embedding tensor satisfying the linear constraint and including the singlet deformation $\Theta$, see e.g. \cite{Danielsson:2013qfa,Dibitetto:2015bia}. A particularly interesting feature is that we here automatically obtain the term
\begin{equation}
 \eta^{MQ} \eta^{NR} \eta^{PS} f_{MNP} f_{QRS} \,,
\end{equation}
which vanishes in truncations of double field theory when the section condition is fulfilled by the background.

\subsection{Reducing the kinetic terms}\label{s:SUGRAKinTerms}

\subsubsection{Scalar kinetic terms}
Consider first the scalar kinetic terms. These were given by
\begin{equation}
 g^{\mu\nu} D_\mu A^a D_\nu A_a \,, \qquad \textrm{ and} \quad \left( g^{\mu\nu} D_\mu B_{u,ab} D_\nu B^u{}_{cd} \right) \epsilon^{abcde} A_e \,.
\end{equation}
Let us begin with the kinetic terms of $A^a$. From the reduction Ansatz we find
\begin{equation}
 \langle D_\mu A^a \rangle = \frac{1}{\sqrt{2}} e^{-4d/5} n^a \left( -\frac{4}{5} \partial_\mu d + A_\mu{}^M f_M \right) \equiv - \frac4{5\sqrt{2}} n^a e^{-4d/5} \D_\mu d \,, \label{eq:SUGRADmud}
\end{equation}
where we defined the gauge-covariant derivative of the dilaton, and thus
\begin{equation}
 \langle g^{\mu\nu} D_\mu A^a D_\nu A_a \rangle = \frac{8}{25} \rho^{-2} e^{4d/5} \bag^{\mu\nu} \D_\mu d \D_\nu d \,. \label{eq:SUGRADMuA2}
\end{equation}

Similarly, for $B_{u,ab}$ we find
\begin{equation}
 \langle D_\mu B_{u,ab} \rangle = \omega_{M,ab} \left( \partial_\mu \left( b_{u}{}^M e^{-2d/5} \right) - A_\mu{}^N b_u{}^P e^{-2d/5} g_{MN}{}^P \right) \equiv \omega_{M,ab} \D_\mu \left( b_u{}^M e^{-2d/5} \right) \,.
\end{equation}
From \eqref{eq:SUGRADmud} one can now read off $\D_\mu b_u{}^M$. We can now calculate
\begin{equation}
 \langle \left( D_\mu B_{u,ab} D_\nu B^u{}_{cd} \right) \epsilon^{abcde} A_e \rangle = 2 \sqrt{2} \D_\mu b_u{}^M \D_\nu b^u{}_M + \frac{24\sqrt{2}}{25} \D_\mu d \D_\nu d \,.
\end{equation}
But from \eqref{eq:bH} we find
\begin{equation}
 \D_\mu \gH^{MN} \D_\nu \gH_{MN} = 8 \D_\mu b_u{}^M \D_\nu b^u{}_M \,,
\end{equation}
and hence
\begin{equation}
 \langle \left( D_\mu B_{u,ab} D_\nu B^u{}_{cd} \right) \epsilon^{abcde} A_e \rangle = \frac{\sqrt{2}}{4} \D_\mu \gH^{MN} \D_\nu \gH_{MN} + \frac{24\sqrt{2}}{25} \D_\mu d \D_\nu d \,. \label{eq:SUGRADmuB2}
\end{equation}

Combining \eqref{eq:SUGRADMuA2} and \eqref{eq:SUGRADmuB2} we find
\begin{equation}
 \begin{split}
  \langle |e| L_{\textrm{SK}} \rangle &= \langle |e| g^{\mu\nu} \left( \sqrt{2} D_\mu B_{u,ab} D_\nu B^u{}_{cd} \epsilon^{abcde} A_e - 56 D_\mu A^a D_\nu A_a \right) \rangle \\
  &= \rho^{5} |\bae| e^{-2d} \left( \frac12 \bag^{\mu\nu} \D_\mu \gH^{MN} \D_\nu \gH_{MN} + 16 \bag^{\mu\nu} \D_\mu d\, \D_\nu d \right) \,.
 \end{split}
\end{equation}
This is the correct kinetic term for the scalars of seven-dimensional half-maximal gauged SUGRA, see e.g. \cite{Dibitetto:2015bia}.

\subsubsection{Gauge kinetic terms and topological term}
Let us first of all consider the reduction of the field strength. We find
\begin{equation}
 \begin{split}
  \langle \Fa_{\mu\nu}{}^{ab} \rangle &= \rho\, \omega_M{}^{ab} F_{\mu\nu}{}^M \,, \\
  \langle \Fb_{\mu\nu\gamma\,a} \rangle &= - 4 \rho^2 n_a H_{\mu\nu\gamma} \,, \\
  \langle \Fc_{\mu\nu\gamma\sigma}{}^a \rangle &= \rho^3 n^a J_{\mu\nu\gamma\sigma} \,,
 \end{split}
\end{equation}
where $F_{\mu\nu}{}^M$, $H_{\mu\nu\gamma}$ and $J_{\mu\nu\gamma\sigma}$ are the reduced field strength of the gauged SUGRA
\begin{equation}
 \begin{split}
  F_{\mu\nu}{}^M & = 2 \partial_{[\mu} A_{\nu]}{}^M - \left[ A_\mu, A_\nu \right]^M - B_{\mu\nu} \left( 2 \xi^M + f^M \right) \,, \\
  H_{\mu\nu\rho} &= 3 \D_{[\mu} B_{\nu\rho]} + 3 \partial_{[\mu} A_\nu{}^M A_{\rho]M} - A_{[\mu}{}^M \left[ A_\nu, A_{\rho]} \right]_M + \frac14 \Theta C_{\mu\nu\rho} \,, \\
  J_{\mu\nu\rho\sigma} &= 4 \D_{[\mu} C_{\nu\rho\sigma]} + \left( \frac32 f^M + \xi^M \right) D_{\mu\nu\rho\sigma\,M} \,.
 \end{split}
\end{equation}
Here $\left[ A_\mu, A_\nu \right]^M$ denotes the Lie bracket of the gauge group defined by the embedding tensor $g_{MNP}$.
\begin{equation}
 \left[ A_\mu, A_\nu \right]^M \equiv g_{NP}{}^M A_\mu{}^N A_\nu{}^P = f_{NP}{}^M A_\mu{}^N A_\nu{}^P - A_{[\mu}{}^M A_{\nu]}{}^N \left(4 \xi_N + f_M \right) \,.
\end{equation}

Let us now consider the kinetic term for the vector fields
\begin{equation}
 L_{\textrm{kin,vectors}} = \frac{1}{8} \Fa_{\mu\nu}{}^{ab} \Fa^{\mu\nu,cd} \left( B_{u,ab} B^u{}_{cd} - B_{u[ab} B^u{}_{cd]} \right) \,. 
\end{equation}
First note that
\begin{equation}
 B_{u[ab} B^u{}_{cd]} = \frac1{\sqrt{2}} \epsilon_{abcde} A^e \,,
\end{equation}
and from \eqref{eq:OmegaIdentities} that
\begin{equation}
 \omega_M{}^{ab} \omega_N{}^{cd} \epsilon_{abcde} = 16 \eta_{MN} n^e \,.
\end{equation}
It is now straightforward to see that
\begin{equation}
 \begin{split}
  \langle |e| L_{\textrm{kin,vectors}} \rangle &= \rho^{5} |\bae| e^{-2d} \bag^{\mu\gamma} \bag^{\nu\sigma} F_{\mu\nu}{}^M F_{\gamma\sigma}{}^N \left( 2 b_{u,M} b^u{}_N - \eta_{MN} \right) \\
  &= - \rho^{5} |\bae| e^{-2d} \bag^{\mu\gamma} \bag^{\nu\sigma} F_{\mu\nu}{}^M F_{\gamma\sigma}{}^N \gH_{MN} \,,
 \end{split}
\end{equation}
which is the correct kinetic term for the vector fields.

Finally, let us reduce the kinetic term for the two-form potentials.
\begin{equation}
 L_{kin,\textrm{2-form}} = -\frac{1}{48} \Fb_{\mu\nu\rho,a} \Fb^{\mu\nu\rho}{}_b A^a A^b \,.
\end{equation}
We find
\begin{equation}
 \langle |e| L_{\textrm{kin,2-form}} \rangle = - \frac{1}{6} \rho^{5} |\bae| e^{-2d} \bag^{\mu\sigma} \bag^{\nu\rho} \bag^{\gamma\lambda} H_{\mu\nu\gamma} H_{\sigma\rho\lambda} \,,
\end{equation}
again reproducing the correct kinetic term for the two-form potentials.

Let us now turn to the topological term. Using \eqref{eq:GaugeAnsatz} we find that the second term of the topological part of the action vanishes, i.e.
\begin{equation}
 \langle \Fa_{\mu_1\mu_2} \bullet \left( \Fb_{\mu_3\ldots\mu_5} \bullet \Fb_{\mu_6\ldots\mu_8} \right) \rangle = 0 \,,
\end{equation}
and we are left with
\begin{equation}
 \langle S_{top} \rangle = \int d^8x d^{10}Y \rho^5 \frac{1}{8\sqrt{6}} \epsilon^{\mu_1\ldots\mu_8} J_{\mu_1\ldots\mu_4} J_{\mu_5\ldots\mu_8} \Theta \,,
\end{equation}
where we used \eqref{eq:SingletFlux}. We see that the singlet deformation $\Theta$ induces a mass-like term for the 3-form $\TC_{\mu\nu\rho}$ \cite{Bergshoeff:2007vb}.

\subsection{Reducing the SUSY variations}\label{s:SUGRASUSY}
Finally, let us use the truncation Ansatz \eqref{eq:SpinorAnsatz} to evaluate the SUSY variations \eqref{eq:N2GravitinoVariation}. For the gravitino variation we find
\begin{equation}
 \begin{split}
  \langle \delta_\epsilon \psi_\mu{}^{\dalpha} \rangle &\sim \D_\mu \epsilon^{\dalpha} - \frac14 \left( e^{-2d} \Theta + \frac{1}{3\sqrt{2}} \gH^{MNP} f_{MNP} \right) \bar{\gamma}_\mu \epsilon^{\dalpha} \\
  & \quad - i \frac{5 \sqrt{2}}{4} e^{4d/5} \left(\sigma_u\right)^\dalpha{}_{\dbeta} b^{u}{}_M \xi^M \bar{\gamma}_\mu \epsilon^{\dbeta} - i e^{2d/5} b_{u,M} F_{\nu\rho}{}^M \left(\sigma^u\right)^{\dalpha}{}_{\dbeta} \bar{\gamma}^{\nu\rho} \bar{\gamma}_\mu \epsilon^{\dbeta} \\
  & \quad + 4 e^{4d/5} H_{\nu\rho\sigma} \bar{\gamma}^{\nu\rho\sigma} \bar{\gamma}_\mu \epsilon^{\dalpha} \,,
 \end{split}
\end{equation}
where $\D_\mu \epsilon^{\dalpha} = \partial_\mu \epsilon^\dalpha - A_\mu{}^M \left( \xi_M - f_M \right) \epsilon^{\dalpha}$ is the gauge-covariant derivative of $\epsilon^{\dalpha}$ and $\bar{\gamma}_\mu = \bae_\mu{}^{\bmu} \gamma_{\bmu}$.

\section{Conclusions}\label{s:Conclusions}
In this paper we showed how to construct seven-dimensional half-maximal consistent truncations of 10- and 11-dimensional supergravity using exceptional field theory. To do this, we began by reformulating the $\SL{5}$ exceptional field theory in a way that is adapted to ${\cal N}=2$ SUSY. In particular, we rewrote the theory by replacing the generalised metric $\gM_{ab}$ with a set of well-defined tensors $\kappa$, $A^a$, $A_a$, $B_{u,ab}$ subject to a compatibility condition, which define the $\SU{2}$-structure. We showed that the existence of these tensors is equivalent to there being two well-defined spinors on the internal space thus ensuring we have ${\cal N}=2$ SUSY. Furthermore, we introduced generalised $\SU{2}$ connections and their intrinsic torsion to rewrite the scalar potential, SUSY variations and kinetic terms of the theory.

A consistent truncation can then be defined by expanding the $\SU{2}$-structure and all other fields of the EFT in terms of a set of sections of the $\left(\mbf{1},\mbf{1}\right)$, the $\left(\mbf{3},\mbf{1}\right)$ and the $\left(\mbf{1},\mbf{3}\right)$-bundles of $\SU{2}_S\times\SU{2}_R \subset \SL{5}$. The number of sections of the $\left(\mbf{3},\mbf{1}\right)$-bundle determines the number of vector multiplets in the gauged SUGRA. These sections were subject to a number of differential constraints, in particular a ``doublet constraint'' which ensured that all $\SU{2}_S$ doublets vanished and hence that we obtained a ${\cal N}=2$ gauged SUGRA. In addition, we had to impose that the sections defining the truncation formed a closed set under the generalised Lie derivative.

We saw that the embedding tensor of the half-maximal gauged SUGRA is now given a geometric definition in terms of the intrinsic $\SU{2}$ torsion of the background on which the truncation is performed. It can thus be written in terms of the generalised Lie derivative of the sections defining the truncation. Exactly as in the case of maximal SUSY, it automatically satisfies the linear constraint of gauged SUGRA and satisfies the quadratic constraint whenever the algebra of generalised diffeomorphisms closes, for example by imposing the section condition. The truncation was shown to be consistent when the embedding tensor is constant.

The framework introduced here can be used to find uplifts of half-maximal seven-dimensional gauged SUGRAs which cannot be obtained by simple truncations of a maximal seven-dimensional gauged SUGRA. These include gauged SUGRAs with non-zero singlet part of the embedding tensor, $\Theta$. A particularly interesting example of these admits a fully stable deSitter vacuum \cite{Dibitetto:2015bia}.

It would also be interesting to generalise the procedure of this paper to lower-dimensions. In this case the number of possible truncations increases because more fluxes are available. For example, it would be nice to study the consistent truncation of IIA on K3 where the full $\ON{20,4}$ duality group should become visible directly in EFT. Another generalisation that is possible in lower dimensions is to consider a larger amount of broken supersymmetry. For example, in four dimensions one could consider ${\cal N}=2$ truncations, corresponding to truncations on generalised $\SU{6}$-structure manifolds, which would include ``exceptional Calabi-Yau''s and their AdS counterparts \cite{Ashmore:2015joa,Ashmore:2016qvs}.

Finally, as we show in \cite{Malek:2016vsh} one can use the set-up introduced here to obtain the heterotic DFT by a reduction of EFT. In this case, the extended space contains a $\SU{2}$-structure manifold, but the coefficients in the truncation Ansatz are still allowed to depend on the extended space, albeit in a restricted fashion. This is reminiscent of the way massive IIA SUGRA can be obtained by a Scherk-Schwarz-like reduction of EFT \cite{Ciceri:2016dmd}. In particular the duality between M-theory on K3 and the heterotic string on $T^3$ arises naturally from this picture.

\section{Acknowledgements}
The author would like to thank Daniel Thompson for initial collaborations on this project. In addition the author thanks David Berman, I\~{n}aki Garc{\'i}a-Etxebarria, Severin L\"ust, Diego Marqu{\'e}s, Carmen Nu{\~n}ez, Erik Plauschinn, Felix Rudolph and Henning Samtleben for helpful discussions. The author would also like to thank IAFE Buenos Aires for hospitality while part of this work was completed. This work is supported by the ERC Advanced Grant ``Strings and Gravity" (Grant No. 320045).

\appendix

\section{Scalar potential in terms of spinors}\label{A:SpinorPot}
As discussed in \cite{Coimbra:2011ky,Coimbra:2012af,Hohm:2014qga} one can write the generalised Ricci scalar in terms of covariant derivatives of spinors, i.e.
\begin{equation}
 \cR \epsilon^i \sim \left( \hat{\nabla}^2 \right) \epsilon^i \,,
\end{equation}
where $\hat{\nabla}$ is the torsion-free $\USp{4}$ connection but without spin connection.

To fix the right-hand side, we follow \cite{Coimbra:2012af} and note that the potential must only involve the determined parts of the torsion-free $\USp{4}$ connection. Also, we know that the right-hand side must be linear in $\epsilon^i$, and hence cannot have any double partial derivatives acting on $\epsilon^i$ (as well as no single partial derivatives acting on $\epsilon^i$). This knowledge is enough to fix the right-hand side. The first observation lets us write
\begin{equation}
 \frac1{16} \cR \epsilon^i = \frac{1}{2}\hat{\nabla}_{jk} \hat{\nabla}^{ki} \epsilon^j - \frac12 \hat{\nabla}_{jk} \hat{\nabla}^{jk} \epsilon^i + \alpha \hat{\nabla}^{ik} \hat{\nabla}_{jk} \epsilon^j \,,
\end{equation}
as we will discuss in \ref{s:DetConnections}, while the second implies that the potential must only make use of the combinations coming from the commutator of two covariant derivatives and their projection onto the $\mathbf{5}$ (since then the second order partial derivatives vanish by the section condition), i.e.
\begin{equation}
 \frac1{16} \cR \epsilon^i = \alpha \left( \frac12 \hat{\nabla}^{ij} \hat{\nabla}_{jk} \epsilon^k - \frac12 \hat{\nabla}_{jk} \hat{\nabla}^{ij} \epsilon^k \right) + \beta \left( \hat{\nabla}^{ik} \hat{\nabla}_{jk} \epsilon^j + \hat{\nabla}_{jk} \hat{\nabla}^{ik} \epsilon^j - \frac12 \hat{\nabla}^{jk} \hat{\nabla}_{jk} \epsilon^i \right) \,.
\end{equation}
These two conditions uniquely fix the Ricci scalar, up to an overall coefficient, to be
\begin{equation}
 \frac{1}{16} \cR \epsilon^i = \frac{1}{2}\hat{\nabla}_{jk} \hat{\nabla}^{ki} \epsilon^j - \frac12 \hat{\nabla}_{jk} \hat{\nabla}^{jk} \epsilon^i + \frac32 \hat{\nabla}^{ik} \hat{\nabla}_{jk} \epsilon^j \,, \label{eq:VSpinors2}
\end{equation}
and in particular the right-hand side is linear in $\epsilon^i$.

\subsection{Determined connections}\label{s:DetConnections}
We must use the covariant derivatives which only depend on the determined part of the torsion-free $\USp{4}$ connection \cite{Coimbra:2011ky,Coimbra:2012af}. There are four different possible combination, depending on whether we act on a spinor in the $\mbf{4}$ or $\mbf{16}$ of $\USp{4}$. Let us denote a generic spinor in the $\mbf{4}$ by $\epsilon$ and a generic spinor in the $\mbf{16}$ by $\chi$ which thus satisfies
\begin{equation}
 \chi^{ij,k} = \chi^{[ij],k} \,, \qquad \chi^{ij,k} \Omega_{ij} = 0 \,, \qquad \chi^{[ij,k]} = 0 \,.
\end{equation}
The unique operators are given by
\begin{equation}
 \nabla \times_4 \epsilon \,, \qquad \qquad \nabla \times_{16} \epsilon \,, \qquad \nabla \times_{4} \chi \,, \qquad \nabla \times_{16} \chi
\end{equation}
with $\times_4$ and $\times_{16}$ being the projectors onto the $\mbf{4}$ and $\mbf{16}$ respectively. In particular, we need
\begin{equation}
 \begin{split}
  \left( \nabla \times_4 \epsilon \right)^i &= \nabla^{ij} \epsilon_j \,, \\
  \left(\nabla \times_{16} \epsilon\right)^{ij,k} &= \nabla^{k[i} \epsilon^{j]} + \frac13 \left( \Omega^{ij} \nabla^{kl} \epsilon_l + \Omega^{k[i} \nabla^{j]l} \epsilon_l \right) \,, \\
  \left( \nabla \times_4 \chi \right)^i &= \nabla_{jk} \chi^{ij,k}
 \end{split}
\end{equation}
Now we can write
\begin{equation}
 V \epsilon^i = \tilde{\alpha} \left( \nabla \times_4 \left( \nabla \times_{16} \epsilon \right) \right)^i + \tilde{\beta} \left( \nabla \times_4 \left( \nabla \times_4 \epsilon \right) \right)^i \,.
\end{equation}

On the other hand, we use the commutator $\left[ \nabla, \nabla \right] \in \mbf{35} \oplus \mbf{10}$ of $\USp{4}$. But since $\mbf{10} \times \mbf{4} \ni \mbf{4}$ and $\mbf{35} \times \mbf{4} \not{\ni} \mbf{4}$ only the $\mbf{10}$ can contribute when acting on $\epsilon^i$. This is given by
\begin{equation}
 \left[ \nabla, \nabla \right]_{10}^{ij} = \nabla^{k(i} \nabla_{k}{}^{j)} \,.
\end{equation}
The other allowed combination involves the projector onto the $\mbf{5}$ since this gives the section condition for the terms involving only partial derivatives. We write
\begin{equation}
 \left( \nabla \times_5 \nabla \right)^{ij} = \nabla^{i}{}_k \nabla^{jk} + \nabla^{jk} \nabla^i{}_k - \frac12 \Omega^{ij} \nabla^{kl} \nabla_{kl} \,,
\end{equation}
and thus combining the two we have
\begin{equation}
 V \epsilon^i = \frac{\alpha}{2} \left( \nabla^{ik} \nabla_{jk} \epsilon^j - \nabla_{jk} \nabla^{ik} \epsilon^j \right) + \frac{\beta}{2} \left( \nabla^{ik} \nabla_{jk} \epsilon^j + \nabla_{jk} \nabla^{ik} \epsilon^j - \frac12 \nabla^{jk} \nabla_{jk} \epsilon^i \right) \,.
\end{equation}
Equating the two allowed expressions gives the unique answer \eqref{eq:PotentialSpinors} (up to overall rescalings of the coupling constant).

\section{SUSY variations of the gravitino}\label{A:SUSY}

We begin with
\begin{equation}
 \begin{split}
  \delta_{\tilde{\epsilon}} \psi_\mu{}^{\dalpha} &= - \frac{1}{\kappa}\theta_i{}^{\dalpha} \delta_{\tilde{\epsilon}} \tilde{\psi}_\mu{}^i \\
  &\sim - \frac{1}{\kappa} \left[ \theta_i{}^{\dalpha} D_\mu \left( \theta^i{}_{\dbeta} \epsilon^\dbeta \right) + \theta_i{}^{\dalpha} \Vi^{a\,ik} \Vi^b{}_{jk} \gamma_\mu \nabla_{ab} \left( \theta^j{}_\dbeta \epsilon^{\dbeta} \right) + \theta^j{}_{\dbeta} \theta_i{}^{\dalpha} \Vi_a{}^{ik} \Vi_{b\,jk} \Fa_{\nu\rho}{}^{ab} \gamma^{\nu\rho} \gamma_\mu \epsilon^\dbeta \right. \\
  & \quad \left. - \Fb_{\nu\rho\sigma,a} \theta^{i\,\dalpha} \Vi^a{}_{ij} \theta^j{}_\dbeta \gamma^{\nu\rho\sigma} \gamma_\mu \epsilon^\dbeta \right] \,.
 \end{split}
\end{equation}
Let us go through this term-by-term.

We use the product rule to write the first term as
\begin{equation}
 \theta_i{}^{\dalpha} D_\mu \left( \theta^i{}_{\dbeta} \epsilon^{\dbeta} \right) = \left( \theta_i{}^{\dalpha} D_\mu \theta^i{}_{\dbeta} \right) \epsilon^{\dbeta} - \kappa D_\mu \epsilon^{\dalpha} \,.
\end{equation}
We can further rewrite $D_\mu \theta^i{}_\dbeta$ in terms of the intrinsic torsion. By definition \eqref{eq:CovExtDeriv}
\begin{equation}
 \begin{split}
  D_\mu \theta^i{}_{\dbeta} &= \partial_\mu \theta^i{}_{\dbeta} - \gL_{A_\mu} \theta^i{}_{\dbeta} \\
  &= \partial_\mu \theta^i{}_{\dbeta} - \left( \gL_{A_\mu}^{\hat{\nabla}} \theta^i{}_{\dbeta} - \frac{1}{20} A_\mu{}^{ab} \tau_{ab} \theta^i{}_{\dbeta} \right) \\
  &= \partial_\mu \theta^i{}_{\dbeta} + \frac1{20} A_\mu{}^{ab} \tau_{ab} \theta^i{}_{\dbeta} - \gL_{A_\mu}^{\hat{\nabla}} \theta^i{}_{\dbeta} \,,
 \end{split}
\end{equation}
where we have introduced an $\SU{2}$ connection $\hat{\nabla}$ and used the definition of the torsion via the generalised Lie derivative \eqref{eq:LieTorsion2}.

The second term can also be rewritten using the intrinsic torsion. We first write it as
\begin{equation}
 \begin{split}
  \theta_i{}^{\dalpha} \Vi^{a\,ik} \Vi^b{}_{jk} \gamma_\mu \nabla_{ab} \left( \theta^j{}_\dbeta \epsilon^{\dbeta} \right) &= - \sqrt{2} \theta_i{}^{\dalpha} \Vi^{ab\,i}{}_j \gamma_\mu \nabla_{ab} \left( \theta^j{}_{\dbeta} \epsilon^{\dbeta} \right) \\
  &= - \sqrt{2} \Vi^{ab\,i}{}_j \gamma_\mu \left( \theta_i{}^{\dalpha} \nabla_{ab} \theta^j{}_{\dbeta} \right) \epsilon^{\dbeta} + \sqrt{2} \Vi^{ab}{}_{ij}  \theta^{i\,\dalpha} \theta^j{}_{\dbeta} \gamma_\mu \nabla_{ab} \epsilon^{\dbeta} \,.
 \end{split}
\end{equation}
We now first use the relationship between $B_{u,ab}$ and $\Vi_{ab}{}^{ij}$ in equation \eqref{eq:ABVi}. Thus we have
\begin{equation}
 \begin{split}
  \theta_i{}^{\dalpha} \Vi^{a\,ik} \Vi^b{}_{jk} \gamma_\mu \nabla_{ab} \left( \theta^j{}_\dbeta \epsilon^{\dbeta} \right) &=  - \sqrt{2} \Vi^{ab\,i}{}_j \gamma_\mu \left( \theta_i{}^{\dalpha} \nabla_{ab} \theta^j{}_{\dbeta} \right) \epsilon^{\dbeta} + \sqrt{2} \Vi^{ab}{}_{ij} \theta^{i\,\dalpha} \theta^j{}_{\dbeta} \gamma_\mu \nabla_{ab} \epsilon^{\dbeta} \\
  &= \frac{1}{\sqrt{2}} \gamma_\mu \left( \theta_i{}^{\dalpha} \nabla^{ij} \theta_{j\,\dbeta} \right) \epsilon^{\dbeta} + \sqrt{2} i V_u{}^{ab} \left(\sigma^u\right)_{\dalpha\dbeta} \gamma_\mu \nabla_{ab} \epsilon^{\dbeta} \,.
 \end{split}
\end{equation}

The first term on the right is proportional to the intrinsic torsion and can thus also be rewritten in terms of the spinor bilinears using \eqref{eq:TorsionSpinors}. To do this we first decompose it into its irreducible representations
\begin{equation}
 \begin{split}
  \frac{1}{\sqrt{2}} \left( \theta_i{}^{\dalpha} \nabla^{ij} \theta_{j}{}^{\dbeta} \right) &= - \frac{1}{2\sqrt{2}} \epsilon^{\dalpha\dbeta} \left( \theta_i{}^{\dgamma} \nabla^{ij} \theta_{j\,\dgamma} \right) + \frac{1}{\sqrt{2}} \left( \theta_i{}^{(\dalpha} \nabla^{ij} \theta_{j}{}^{\dbeta)} \right) \\
  &= - \frac{\kappa}{2} \left( S + \frac{T}{\sqrt{2}} \right) + \frac{\kappa}{4} S^{\dalpha\dbeta} \,.
 \end{split}
\end{equation}

Hence we have that
\begin{equation}
 \begin{split}
  \theta_i{}^{\dalpha} \Vi^{a\,ik} \Vi^b{}_{jk} \gamma_\mu \nabla_{ab} \left( \theta^j{}_\dbeta \epsilon^{\dbeta} \right) &= \frac{\kappa}{2} \left( S + \frac{T}{\sqrt{2}} \right) \gamma_\mu \epsilon^{\dalpha} + \frac{\kappa}{4} S^{\dalpha}{}_{\dbeta} \gamma_\mu \epsilon^{\dbeta} \\
  & \quad + \sqrt{2} i V_u{}^{ab} \left(\sigma^u\right)^{\dalpha}{}_{\dbeta} \gamma_\mu \nabla_{ab} \epsilon^{\dbeta} \,.
 \end{split}
\end{equation}

The third and fourth term follow similarly and we get
\begin{equation}
 \begin{split}
  \theta^j{}_{\dbeta} \theta_i{}^{\dalpha} \Vi_a{}^{ik} \Vi_{b\,jk} \Fa_{\nu\rho}{}^{ab} \gamma^{\nu\rho} \gamma_\mu \epsilon^\dbeta &= i \sqrt{2} B_{u,ab} \Fa_{\nu\rho}{}^{ab} \left(\sigma^u\right)^{\dalpha}{}_{\dbeta} \gamma^{\nu\rho} \gamma_\mu \epsilon^{\dbeta} \,, \\
  - \Fb_{\nu\rho\sigma,a} \theta^{i\,\dalpha} \Vi^a{}_{ij} \theta^j{}_\dbeta \gamma^{\nu\rho\sigma} \gamma_\mu \epsilon^\dbeta &= \frac12 \Fb_{\nu\rho\sigma,a} A^a \gamma^{\nu\rho\sigma} \gamma_\mu \epsilon^{\dalpha} \,.  
 \end{split}
\end{equation}

Putting everything together we find the $\left({\cal N}=2\right)$-like gravitino variation.
\begin{equation}
 \begin{split}
  \delta_\epsilon \psi_\mu{}^{\dalpha} &\sim D_\mu \epsilon^{\dalpha} - \frac1{\kappa} \left(\theta_i{}^{\dalpha} \partial_\mu \theta^i{}_{\dbeta}\right) \epsilon^{\dbeta} + \frac1{20} A_\mu{}^{ab} \tau_{ab} \epsilon^{\dalpha} + \frac1\kappa \theta_i{}^{\dalpha} \left(\gL_{A_\mu}^{\hat{\nabla}} \theta^i{}_{\dbeta}\right) \epsilon^{\dbeta} \\
  & \quad - \frac\kappa2 \left( S + \frac{T}{\sqrt{2}} \right) \gamma_\mu \epsilon^{\dalpha} - \frac{\kappa}{4} S^{\dalpha}{}_{\dbeta} \gamma_\mu \epsilon^{\dbeta} \\
  & \quad - i\sqrt{2} V_u{}^{ab} \left(\sigma^u\right)^{\dalpha}{}_{\dbeta} \gamma_\mu \nabla_{ab} \epsilon^{\dbeta} - i \sqrt{2} B_{u,ab} \Fa_{\nu\rho}{}^{ab} \left(\sigma^u\right)^{\dalpha}{}_{\dbeta} \gamma^{\nu\rho} \gamma_\mu \epsilon^{\dbeta} \\
  & \quad - \Fb_{\nu\rho\sigma,a} A^a \gamma^{\nu\rho\sigma} \gamma_\mu \epsilon^{\dalpha} \,.
 \end{split}
\end{equation}

\bibliographystyle{JHEP}
\bibliography{NewBib}

\end{document}